\documentclass[prl,twocolumn,floatfix,superscriptaddress]{revtex4-1}
\usepackage{dcolumn,amsmath}
\usepackage{graphicx}
\usepackage{bm}
\usepackage{hyperref} 
\newcommand{\PSen}{PS{\em e}N}
\newcommand{\Eeff}{\ensuremath{E_{\rm eff}}}
\newcommand{\eEDM}{{\em e}EDM}
\newcommand{\ecm}{\ensuremath{e {\cdotp} {\rm cm}}}
\newcommand{\cm}{\ensuremath{{\rm cm}}$^{-1}$}
\newcommand{\de}{d_\mathrm{e}}
\begin{document}
%  \title{Theoretical study of ThO for the electron electric dipole moment search II}
   \title{Theoretical study of thorium monoxide for the electron electric dipole moment search, II: Electronic properties of $H^3\Delta_1$ in ThO}

\author{L.V.\ Skripnikov}\email{leonidos239@gmail.com}
\author{A.V.\ Titov}
\homepage{http://www.qchem.pnpi.spb.ru}
  \affiliation{B.P.~Konstantinov Petersburg Nuclear Physics Institute, Gatchina, Leningrad district 188300, Russia}
\affiliation{Dept.\ of Physics, Saint Petersburg State University, Saint Petersburg, Petrodvoretz 198504, Russia}
\date{\today}

\begin{abstract}

Recently an improved limits on the electron electric dipole moment, \eEDM, and dimensionless constant, $k_{T,P}$, characterizing the strength of the  T,P-odd pseudoscalar$-$scalar electron$-$nucleus neutral current interaction in the $H^3\Delta_1$ state of ThO molecule were obtained by ACME collaboration [Science 343, 269 (2014)]. The interpretation of the experiment in terms of  fundamental quantities \eEDM\ and $k_{T,P}$ is based on the results of theoretical study of appropriate ThO characteristics, the effective electric field acting on electron, \Eeff, and a parameter of the T,P-odd pseudoscalar$-$scalar interaction, $W_{T,P}$, given in [J.Chem.Phys.\ 139, 221103 (2013)] by St.Petersburg group. To reduce the uncertainties of the given limits we report improved calculations of the molecular state$-$specific quantities \Eeff, 81.5~GV/cm, and $W_{T,P}$, 112~kHz, with the uncertainty within 7\% of the magnitudes. Thus, the values recommended to use for the upper limits of the quantities are 75.8~GV/cm and 104~kHz, correspondingly. The hyperfine structure constant, molecule-frame dipole moment of the $H^3\Delta_1$ state and $H^3\Delta_1\to X^1\Sigma^+$ transition energy which, in general, can serve as a measure of reliability of the obtained \Eeff\ and $W_{T,P}$ values are also calculated. Besides we report the first calculation of g-factor for the $H^3\Delta_1$ state of ThO. The results are compared to the earlier experimental and theoretical studies, and a detailed analysis of uncertainties of the calculations is given.
\end{abstract}

\maketitle

%========================================================================
\section{Introduction.}

The search for a permanent electric dipole moment (\eEDM) of electron is one of the most intriguing fundamental problems of modern physics. A nonzero value of \eEDM\ implies manifestation of interactions which are not symmetric with respect to both time (T) and spatial (P) inversions (T,P-odd interactions). The observation of \eEDM\ at the level significantly greater than $10^{-38}\ \ecm$ would indicate the presence of a ``new physics'' beyond the Standard model (see review \cite{Commins:98, Ginges:04, Khriplovich:11, Chupp:14gka} and references). Popular extensions of the Standard model predict the magnitude of the \eEDM\ at the level of  $10^{-26}-10^{-29}\ \ecm$ \cite{Commins:98}.
  
Several decades ago it was realized \cite{Sandars:64, Sandars:65} that very prospective experiments towards the search of violation of fundamental symmetries could be performed on heavy atoms. It was noted soon that even more promising experiments can be done on molecules containing heavy elements \cite{Sandars:67, Onischuk:67}, and  first experimental studies of ``nuclear-induced'' T,P-odd interactions on TlF beam were performed in Oxford \cite{Sandars:67, Harrison:69a, Hinds:76, Hinds:80b}. Systematic theoretical investigations of heavy-atom diatomic molecules with nonzero total electronic momenta 
%   unpaired electrons
were started by the Novosibirsk and St.Petersburg groups some time later \cite{Labzowsky:78, Sushkov:78, Gorshkov:79, Sushkov:84, Flambaum:85b, Kozlov:87}. In these studies a unique enhancement mechanism of T,P-odd electron-nuclear and ``electron-induced'' effects based on mixing close $\Omega-$doublet sublevels of opposite parity was exploited. Soon, the experimental search of \eEDM\ on the YbF molecular beam was started at Yale University by Hinds with colleagues \cite{Sauer:94}, however, only in 2011 the YbF experiments~\cite{Hudson:11a} gave more rigid limit on \eEDM\ ($1.05 \cdotp 10^{-27}\ \ecm$) than the atomic Tl beam~\cite{Regan:02} measurements ($1.6 \cdotp 10^{-27}\ \ecm$). The latest limit on the electron EDM, $|\de|<8.7\times 10^{-29}$ \ecm\ (90\% confidence), was set with the molecular beam of thorium monoxide (ThO) molecules in the metastable electronic $H^3\Delta_1$ state \cite{ACME:14a} exploiting unique advantages of this ThO state in the \eEDM\ enhancement and suppressing the experimental uncertainties. Besides, the experiment provided also a limit on the T,P-odd pseudoscalar$-$scalar electron$-$nucleus (\PSen) neutral currents interaction \cite{ACME:14a} dimensionless constant, $k_{T,P} < 5.9\times 10^{-9}$ (90\% confidence). It was estimated in Ref.~\cite{Pospelov:14} within the Standard model that the interaction can induce even greater T,P-odd effect simulating the \eEDM. Recently it was also proposed in Refs.~\cite{FDK14, Skripnikov:14a} to use $^{229}$ThO to study T,P-odd interaction of the quadrupole magnetic moment of the $^{229}$Th nucleus with electrons. Besides, a new series of experiments is under preparation on the thorium monofluoride cation, ThF$^+$, (which is isoelectronic to ThO) by E.~Cornell group \cite{Leanhardt:2011}.

Actually, in the experiment \cite{ACME:14a} the limits on interaction energy of the electron EDM with the effective electric field (\Eeff, see text after Eq.~(\ref{matrelem})) on unpaired electrons in ThO as well as the \PSen\ interaction energy were obtained. To get the limit on the \eEDM\ and $k_{T,P}$ the values of \Eeff=84~GV/cm and $W_{T,P}{=}116$~kHz (a parameter characterizing the molecular state$-$specific part of the later interaction, see Eq.~(\ref{WTP})) were taken from Ref.~\cite{Skripnikov:13c}. The theoretical uncertainties of \Eeff\ and $W_{T,P}$ were conservatively estimated in \cite{Skripnikov:13c} as 15\%. Nowadays it is anticipated that sensitivity of the ThO experiment can be considerably improved by as much as $\sim\!2$ orders of magnitude \cite{ACME_Improvements}. In particular, in Ref.~\cite{Petrov:14} it was proposed to use the second rotational level (rather than the first one exploited in \cite{ACME:14a}) of the $H^3\Delta_1$ state to suppress systematic errors related to magnetic fields since even a complete coincidence of g-factors
%   (almost zero g-factors difference)
can be achieved for the $\Omega$ doublet levels in this case in a certain external electric field (zero (almost zero) g-factors difference can lead to efficient suppression of main uncertainties in ThO experiment).
% due to magnetic field gradient, stray etc.\ magnetic fields) systematic magnetic 
Therefore, an improved accuracy of the effective electric field and other parameters which cannot be measured but required to interpret the experimental results in terms of fundamental quantities is of interest.
Recently, some four-component (Dirac-Coulomb) multireference configuration interaction calculations (MRCI) of \Eeff\ in the $H^3\Delta_1$ state of ThO molecule were performed in Ref.~\cite{Fleig:14} with a very small declared uncertainty for \Eeff, 3\%.
% that looks rather surprising for the actinide-containing oxides.

Improvement of computational technologies to attain a high accuracy for different properties of actinide compounds is on the agenda for many application concerning both fundamental and practical purposes. This is also a motivation for us in context of studying the ThO properties,
% required to search for the T,P-odd effects,
since the goals of the present research are:\\
 (i) analysis of applicability of the most popular correlation approaches, the coupled-clusters (CC) one, used earlier in Ref.~\cite{Skripnikov:13c}, and MRCI, applied in Ref.~\cite{Fleig:14}, to get the best to-date values of \Eeff\ and $W_{T,P}$;\\
 (ii) discussion of theoretical uncertainties of the MRCI and coupled-clusters calculations;\\
 (iii) developing and applying a combined computational scheme for
% performing
a new series of precise studies of different properties of ThO with reliable estimation of their theoretical uncertainties;\\
(iv) calculation of g-factor for the $H^3\Delta_1$ state of ThO to
%   test the possibility of
    check an ability of the methods for its accurate theoretical study.

%-----------------------------
    \section{Theoretical details}
%   \section{Theoretical background}

In the present paper we have considered the transition energy between the ground $^1\Sigma^+$ and the metastable $H ^3\Delta_1$ state ($T_e$), molecule-frame dipole moment ($d$), effective electric field (\Eeff), the molecule-specific parameter of the \PSen\ interaction ($W_{T,P}$), hyperfine structure constant (A$_{||}$)
as well as g-factor
 for the $H ^3\Delta_1$ state of ThO. The effective Hamiltonian parameters \Eeff, $W_{T,P}$, A$_{||}$ are examples of so-called ``atom in a compound'' (AiC) characteristics \cite{Titov:14}. The quantities are the mean values of the operators heavily concentrated in the atomic Th core and they are mainly sensitive to variation of core-region spin densities of the valence and outer-core electrons.

As shown in Refs.~\cite{Titov:06amin, Petrov:02} accurate and efficient computation of such characteristics can be performed by the two-step approach \cite{Titov:06amin, Petrov:02} utilizing the generalized relativistic effective core potential (GRECP) method \cite{Titov:99, Mosyagin:10a}. At the first (molecular) step, the GRECP is used to exclude the inner-core electrons from a correlation calculation and obtain an accurate description of the valence part of the wave function by an economical way. Thus, the computational cost of the relativistic molecular calculation is dramatically reduced compared to a conventional all-electron four-component case (in particular, due to ability to perform scalar-relativistic corrections, see below).
It should be noted that the GRECP operator also allows one to take account of the Breit interaction very effectively \cite{Petrov:04b, Mosyagin:06amin}. Second, a nonvariational restoration procedure is employed \cite{Titov:06amin} to recover the valence wave function in the inner core region of a heavy atom. The procedure is based on a proportionality of valence and virtual spinors in the inner-core regions of heavy atoms. To perform the restoration one generates {\it equivalent} basis sets of one-center four-component spinors
$$
%   \left\{
  \left( \begin{array}{c} f_{nlj}(r)\theta_{ljm} \\
     g_{nlj}(r)\theta_{2j{-}l,jm} \\ \end{array} \right)
%  \right\}
$$
 and smoothed two-component pseudospinors
$$
    \tilde f_{nlj}(r)\theta_{ljm}
$$
in all-electron finite-difference Dirac-Fock-Breit and GRECP/\,self-consistent field calculations (employing the $jj-$coupling scheme) of {the same} configurations of a considered atom and its ions \cite{HFDB, Bratzev:77, HFJ, Tupitsyn:95}. These sets, describing mainly the given atomic core region, are generated independently of the basis set exploited in the molecular GRECP calculations. A first-order reduced density matrix obtained at the first step is reexpanded into the basis set of smoothed two-component pseudospinors. Replacing these pseudospinors by equivalent four-component spinors one obtains the true four-component density matrix. Taking trace of the product of the density matrix with the matrix form of an operator describing a given property one obtains the expectation value of the property. Note that the numerical form of four-component spinors is used which allows one to get a correct behavior of the wavefunction in the core region of a given heavy atom.

To obtain \Eeff\ one can evaluate an expectation value of the a T,P-odd operator (discussed in Refs.~\cite{Kozlov:87, Kozlov:95, Titov:06amin}):
\begin{equation}
\label{matrelem}
W_d = \frac{1}{\Omega}
\langle \Psi|\sum_i\frac{H_d(i)}{d_e}|\Psi
\rangle,
\end{equation}
where $d_e$ is the value of \eEDM, $\Psi$ is the wave function of the considered state of ThO, and
$\Omega= \langle\Psi|\bm{J}\cdot\bm{n}|\Psi\rangle$,
$\bm{J}$ is the total electronic momentum, $\bm{n}$ is the unit vector along the molecular axis directed from Th to O
($\Omega{=}1$ for the considered $H^3\Delta_1$ state of ThO),
\begin{eqnarray}
  H_d=2d_e
  \left(\begin{array}{cc}
  0 & 0 \\
  0 & \bm{\sigma E} \\
  \end{array}\right)\ ,
 \label{Wd}
\end{eqnarray}
$\bm{E}$ is the inner molecular electric field, and $\bm{\sigma}$ are the Pauli matrices. In these designations $E_{\rm eff}=W_d|\Omega|$.

The T,P-odd pseudoscalar$-$scalar electron$-$nucleus interaction with a characteristic dimensionless constant $k_{T,P}$ is given by the following operator (see \cite{Hunter:91}):
\begin{eqnarray}
  H_{T,P}=i\frac{G_F}{\sqrt{2}}Zk_{T,P}\gamma_0\gamma_5n(\textbf{r}),
 \label{Htp}
\end{eqnarray}
where $G_F$ is the Fermi-coupling constant, $\gamma_0$ and $\gamma_5$ are the Dirac matrixes and $n(\textbf{r})$ is the nuclear density normalized to unity. To extract the fundamental $k_{T,P}$ constant from an experiment one
needs to know the factor $W_{T,P}$ that is determined by the electronic structure of a studied molecular state
on a given nucleus:
\begin{equation}
\label{WTP}
W_{T,P} = \frac{1}{\Omega}
\langle \Psi|\sum_i\frac{H_{T,P}(i)}{k_{T,P}}|\Psi
\rangle\ .
\end{equation}
The \Eeff\ and $W_{T,P}$ parameters cannot be measured and have to be obtained from a molecular electronic structure calculation.

The accuracy of calculated values of \Eeff\ and $W_{T,P}$ can be estimated only indirectly. For this one can calculate the mean value of an operator (within the same approximation for the wave function) which have comparable to \Eeff\ and $W_{T,P}$ sensitivity to different variations of wave function but, in contrast, the operator should correspond to the property which can be measured. Similar to \Eeff\ and $W_{T,P}$ these parameters should be sensitive to a change of densities of the \textit{valence} electrons in atomic core(s).
The hyperfine structure constant, $A_{||}$, is traditionally used as such a parameter (e.g., see \cite{Kozlov:97}).
To obtain $A_{||}$ on Th in the ThO molecule theoretically the following matrix element can be evaluated:
\begin{equation}
 \label{Apar}
A_{||}=\frac{\mu_{\rm Th}}{I\Omega}
   \langle
   \Psi|\sum_i\left(\frac{\bm{\alpha}_i\times
\bm{r}_i}{r_i^3}\right)
_z|\Psi
   \rangle, \\
\end{equation}
where $\mu_{\rm Th}$ is magnetic moment of an isotope of Th nucleus having spin $I$,
$ \bm{\alpha}=
  \left(\begin{array}{cc}
  0 & \bm{\sigma} \\
  \bm{\sigma} & 0 \\
  \end{array}\right).
$
To exclude uncertainty of $A_{||}$ caused by the experimental uncertainty of $\mu_{\rm Th}$ \cite{Safronova:13} the magnetic moment independent units $\mu_{\rm Th} / \mu_{\rm N} \cdot$MHz (where $\mu_{\rm N}$ is the nuclear magneton) are used in the present paper. Note that because of absence of experimental data on $A_{||}$ it is not possible to check the calculated value of $A_{||}$ now. As in our previous study of \Eeff\ in ThO \cite{Skripnikov:13c} we also calculated the excitation energy, $T_e$, from the ground $X^1\Sigma$ state to that of our interest, $H^3\Delta_1$, as well as the molecule-frame dipole moment of the $H^3\Delta_1$ state. Both the parameters cannot be considered as a good check for \Eeff, $W_{T,P}$ and other studied AiC parameters. Nevertheless, the accuracy of somewhat related information to the properties of interest like the molecule-frame dipole moment is more
relevant than that of evaluated $T_e$ because it checks the quality of the wave-function obtained for the state of interest
%   (mainly for valence electrons) and not a differential characteristic that loses some important state$-$specific information even for valence electrons.
and mainly for the valence electrons.
In turn, the differential characteristics like transition energies miss some important state$-$specific information for outermost core and even for valence electrons. It was earlier shown on different molecules (see \cite{Titov:06amin} and references) that the outer core spin relaxation/correlation effects can contribute to the T,P-odd effects considerably, up to about 50\% from contribution of the valence electrons whereas these effects can be negligible for the energetic properties.

In addition we have calculated the value of g-factor for the $H^3\Delta_1$ state. It is defined as 
\begin{eqnarray}
 \label{Gpar}
  G_{\parallel} &=&\frac{1}{\Omega} \langle H^3\Delta_1 |\hat{L}^e_{\hat{n}} - g_{S} \hat{S}^e_{\hat{n}} |H^3\Delta_1 \rangle,  
\end{eqnarray}
where ${\vec{L}}^e$ and ${\vec{S}}^e$ are the electronic orbital and electronic spin momentum operators, respectively; $g_{S} = -2.0023$ is a free$-$electron $g$-factor. The value of $G_{\parallel}$ was extracted from the experimental datum \cite{Kirilov:13} in Ref.~\cite{Petrov:14}. Note that the value of $G_{\parallel}$ is close to zero for the $H^3\Delta_1$ state (and equal to zero in the scalar-relativistic approaches where
%   quantum-electrodynamics correction
radiation corrections to free-electron g-factor are also omitted).
It is a more computationally sensitive parameter (in the relativistic case) than the other ones considered here to the quality of the wave function (since high-order interference contributions between the spin-orbit and electron correlation  effects are important here).

The ThO spectroscopy was earlier studied both experimentally and theoretically in a number of papers \cite{Wang:11c, Buchachenko:10, Paulovic:03, Meyer:08}. First estimate of \Eeff\ in the $^3\Delta_1$ state of ThO was obtained in Ref.~\cite{Meyer:08}. The experimental measurements of molecule-frame dipole moments of the ground $X ^1\Sigma^+$ and excited $E$ states were performed in Ref.~\cite{Wang:11c}. In Ref.~\cite{Buchachenko:10} a theoretical study of the potential energy curve and electric properties of the ground state of ThO was done. A series of relativistic calculations of the spectroscopic parameters of the ground and excited states were performed in Ref.~\cite{Paulovic:03}.

{\sc dirac12} code \cite{DIRAC12} was used to make the two-component Hartree-Fock calculations  and, correspondingly, the required one- and two-electron Hamiltonian matrix elements. {\sc cfour} code \cite{CFOUR,Gauss:91,Gauss:93} was used to perform scalar-relativistic Hartree-Fock calculations as well as the coupled-clusters ones with single, double (and  non-iterative triple cluster amplitudes). {\sc mrcc} code \cite{MRCC2013} was used to perform one- and two-component coupled-clusters calculations (with up to quadruple cluster amplitudes included) and all of the presented MRCI calculations. The nonvariational restoration code developed in Refs.~\cite{Skripnikov:13b, Skripnikov:13c, Skripnikov:11a} and interfaced to these program packages was used to restore the four-component electronic structure near the Th nucleus.

%------------------------------
\section{Computational details}

The $1s-4f$ inner-core electrons of Th were excluded from molecular correlation calculations using the valence (semi-local) version of GRECP \cite{Mosyagin:10a} operator. Thus, the outermore 38 electrons ($5s^2 5p^6 5d^{10} 6s^2 6p^6 6d^2 7s^2 $ (Th) and $1s^2 2s^2 2p^4$(O)) were treated explicitly. 

A number of basis sets were used in the present paper.
(i) CBas  which consists of $6s$, $5p$, $3d$, $3f$ contracted Gaussians for Th and $4s$, $3p$ for oxygen.
(ii) CBasSO  which consists of $6s$, $8p$, $5d$, $3f$ Gaussians for Th and $4s$, $3p$ for oxygen.
(iii) MBas  which consists of $30s$, $8p$, $10d$, $4f$, $4g$ and $1h$ Gaussians for Th and can be written as (30,20,10,11,4,1)/[30,8,10,4,4,1], so the only $p$ and $f$ Gaussians are contracted. This basis is the extension of the basis used in Ref.~\cite{Skripnikov:13c}. For oxygen the aug-ccpVQZ basis set \cite{Kendall:92} with removed two g-type basis functions was employed, i.e., we used the (13,7,4,3)/[6,5,4,3] basis set.
(iv) LBas:  [22,17,15,14,10,10,5]\footnote{The number of $s$ basis functions was reduced  because of a linear dependence problem, however, it did not influence the accuracy of the calculated parameters} for thorium and aug-ccpCVQZ basis set \cite{Kendall:92} with removed $g$-type basis functions, (16,10,6,4)/[9,8,6,4] for oxygen.

To generate compact basis set, CBas, we used the procedure developed in Ref.~\citep{Skripnikov:13a}. It is similar to that employed for generating atomic natural basis sets \cite{Almlof:87}.
%   The atomic blocks from the density matrix obtained in the scalar-relativistic coupled-clusters method with single, double and non-iterative triple cluster amplitudes, CCSD(T), calculation of the $^3\Delta$ state of ThO with the LBas basis set were diagonalized to yield the atomic natural-like basis set.
    The atomic blocks from the density matrix obtained in the scalar-relativistic calculation of the $^3\Delta$ state of ThO (within the coupled-clusters method with single, double and non-iterative triple cluster amplitudes, CCSD(T), and with the LBas basis set) were diagonalized to yield the atomic natural-like basis set.
The functions with the largest occupation numbers were selected from these natural basis functions. The results obtained with the given basis set approximately (within 2-3\%) reproduce those with Mbas basis set. CBasSO, which was used in the two-component calculations was obtained from CBas by addition of three $p$-type orbitals and 2 $d$-type orbitals. It is required for accurate reproducing the essentially different radial parts of the $5p_{1/2}$ and $5p_{3/2}$, $6p_{1/2}$ and $6p_{3/2}$, $7p_{1/2}$ and $7p_{3/2}$, as well as $5d_{3/2}$ and $5d_{5/2}$, $6d_{3/2}$ and $6d_{5/2}$ spinors of Th.

Within the (G)RECP approach it is possible to exclude
   naturally
the spin-orbit effects for valence electrons only and perform the scalar-relativistic calculations \cite{Mosyagin:10a}. This leads to considerable computational savings and allows one to use larger basis sets with respect to two-component study when exploiting the same computational resources. We used this feature to perform a series of calculations towards choosing an optimal method of accounting for electron correlation. Besides it was used to calculate the correction on the basis set enlargement for the transition energy between the ground $X ^1\Sigma^+$ and excited $H^3\Delta_1$ states as well as the molecule-frame dipole moment, \Eeff, $W_{T,P}$ and $A_{||}$ constants for the $^3\Delta_1$ state of ThO.

The experimental equilibrium internuclear distances \cite{Huber:79,Edvinsson:84} (3.478 a.u.\ for the $X ^1\Sigma^+$ state and 3.511 a.u.\ for the $H^3\Delta_1$ state) were used in the present calculations. It was shown in our previous paper \cite{Skripnikov:13c} that the calculated equilibrium internuclear distances as well as harmonic frequencies are very close to the experimental values \cite{Huber:79,Edvinsson:84}.

%--------------------------------------------
% \section{Comparison of correlation methods}
% \section{Correlation contributions}
% \section{Selection of method to account for electron correlation}
% \section{Correlation contributions and estimates of computational uncertainties}
% \section{Analysis of correlation methods and estimates of computational uncertainties}
  \section{Analysis of uncertainties of correlation methods}

It is well-known that the MRCI methods allow one to reproduce the wave function in the valence region (describing static and nondynamic correlation effects) more reliably in general (with number of correlated electrons up to ${\sim}20$) than the single-reference coupled-clusters ones in complicated cases (such as (quasi)degeneration of valence levels, dense excitation spectra, few open valence shells, etc.). In the latter case large cluster amplitudes appear in the single-reference coupled-cluster expansion which can lead to unstable results, e.g., see \cite{Bartlett:95}. However, it is not less known that the dynamic correlation effects (with explicit treatment of outer core shells, etc.) are much better described by the
    coupled-clusters
approaches. When both nondynamic and dynamic correlations are important it is often reasonable to apply some combining schemes like that has been first applied at the correlation calculations of T,P-parity nonconservation effects in the PbO molecule \cite{Petrov:05a, Isaev:04}.

The situation with the $H^3\Delta_1$ state in ThO deserves some particular consideration since it is declared in \cite{Fleig:14} that the MRCI method is required to obtain accurate results for \Eeff\ and other parameters of ThO in the $^3\Delta_1$ state. To analyze the situation in detail we have performed a number of calculations using different MRCI and CC methods and the results collected in table~\ref{TCorr} are discussed below. The calculations were performed within the CBas basis set using the one-component (scalar-relativistic) approaches. Note that one should not directly compare the results from Ref.~\cite{Fleig:14} with the results given in the table \ref{TCorr}, as a relatively small basis set is used in the calculations and the spin-orbit effects for valence electrons are omitted. However, on the basis of these calculations one can estimate the importance of different types of excitations including high-order ones which are not accessible in the two-component case due to limitation of computer resources. Only using the results of calculations which include higher-order excitations in the wave-function one can estimate more or less reliably the computational uncertainties of 
%   the methods with lower-order excitations.
   ``standard'' approaches (exploiting the lower-order excitations only).

The used spin-orbitals were obtained within the restricted open-shell Hartree-Fock (ROHF) method for the $^3\Delta$ state of ThO. In the formal ionic model this state corresponds to the $[\dots] 5s^2 5p^6 5d^{10} 6s^2 6p^6 7s^1 6d^1$ effective electron configurations for Th and $1s^2 2s^2 2p^6$ for O. In the calculations, the $5s^2 5p^6 5d^{10}$ shells of Th as well as $1s^2$ shell of O were excluded from the correlation treatment. The explicitly correlated 18 electrons were divided into three groups originated on atomic states of Th and O:
(i) Closed-shell valence orbitals, $6s^2 6p^6$(Th) and $2s^2 2p^6$(O), i.e., 16 spin-orbitals. The orbitals will be denoted below as ``o''.
(ii) Open-shell spin-orbitals, which qualitatively may be considered as $7s$ and 6d$_\delta$ of Th, i.e., six spin-orbitals (note that only one of four 6d$_\delta$ spin-orbitals of Th is occupied in the reference determinant). The orbitals will be denoted below as ``v''.
(iii) All the other (virtual) orbitals.

\begin{table*}[!h]
\caption{
The correlation contributions to the effective electric field (\Eeff) and hyperfine structure constant (A$_{||}$) of the $H ^3\Delta_1$ state of ThO molecule in different 18-electron configuration interaction and coupled-clusters calculations relative to 2e-CISD. 
The results of lines 2--6 are given in the order of improving the level of accounting for the electron correlation. Considered excitations types are: ``o'' means excitations from closed valence space to open-shell and virtual valence space; ``v'' means excitations from open-shell valence space to virtual space; 1v, 2o,  etc. stand for one-fold v-type excitations, two-fold o-type, etc. excitations, respectively; 1o1v stands for simultaneous one-fold o-type and one-fold v-type excitations. 
In the case of single-reference calculations the open-shell space includes two one-electron functions occupied in the leading determinant, otherwise, the open-shell space includes six one-electron functions (3 Kramers pairs): s, 6d$_\delta$ (Th).
``+''/``-'' means that excitations of a given type are treated/omitted by a given method; ``$\pm$''  means that excitations of a given type are  treated only partly within a given method; ``(+)'' means that excitations of a given type are treated only by disconnected diagrams (for coupled-clusters methods).
``OS'' means orbitals from the open-shell space.
}
\label{TCorr}
\begin{tabular}{ l l l l l l l l l   l  r  r  r  r r}
\hline\hline
 ~  &         ~      &    ~   &       ~    &&&&&&  &\multicolumn{2}{c}Fleig and Nayak \cite{Fleig:14}&   \multicolumn{2}{l}  ~CBas basis set ,  &     \\
 ~  &         ~      &    ~   &       ~    &&&&&&  & 4-comp. &   ~ & \multicolumn{2}{l} ~1-comp.   & \\ 
\hline
 \# & \#  & Reference & \multicolumn{6}{c}{Excitation types}                        & Method & \Eeff,~~            & A$_{||}$,~~~ & \Eeff,~~~& A$_{||}$,~~~ & Energy,\\
    &  of & space& 1v,        &1o,  & 1o2v,&2o2v& 3o,&4o &       &   GV/cm & ~$\frac{\mu_{\rm Th}}{\mu_{\rm N}}\cdot$MHz     & ~GV/cm & ~$\frac{\mu_{\rm Th}}{\mu_{\rm N}}\cdot$MHz   & Hartree      \\

    &  act.           &      & 2v         &2o,  & 2o1v,&     &3o1v   &      &    &    &    &  &  &       \\     
    & els.           &      &            &1o1v&       &     &        &     &       &    &    &  &  &       \\

\hline\hline  
% 1  &  2           & Single ref.\,$^a$  &+   &-  & - &- &- &- &-  & CISD   & (68.5) 0.0         &(-2809) 0& (59.4\,$^b$) 0.0  &(-3001) 0 & 0      \\
 1  &  2           & Single ref.  &+   &-  & - &- &- &-   & CISD   &        0.0         &        0&              0.0  &        0 & 0      \\
    &              &$^a$               &    &   &   &  &  &    &        & (68.5)             &(-2809)  & (59.4\,$^b$)      &(-3001)   &        \\ 
 \hline
 2  &  18          &3 OS (s, 6d$_\delta$)      &+    & + &-&- &- &-   & MR(3)-CISD & 12.5                  & -62       & 12.5  & -58 & -0.329 \\ 
 3  &  18          &3 OS+9 virt.&+    & + &$\pm$&$\pm$ &- &-  & MR(12)-CISD $^c$  & 6.7        & -167      &       &     &        \\    
 4  &  18          &3 OS+all virt. &+ & + &+&+ &- &-  &  MR(${\infty}$)-CISD       &     &       &  7.8  & -129& -0.335 \\   
 5  &  18          &3 OS+all virt. &+ & + &+&+ &+ &-  &  (MR(${\infty}$)-CISDT)$_4$&     &       &  11.3 & -129& -0.356 \\    
 6  &  18            & Single ref.    &+ & + &+&+ &+ &+  & CISDTQ                   &     &       &  12.1 & -83 & -0.381 \\   
\hline  
 7  &  18            &3 OS+all occ.  &&&&&&& MR$_3^{+T}$-CISD$^f$            &12.3$^{d,e}$ &  -36$^{d,e}$    &       &     &        \\        
 8  &  18            &3 OS+all occ.  &&&&&&& MR$_3^{+TQ}$-CISD$^g$     &     &        &  13.0 & -49 & -0.331 \\      
 9  &  18          &3 OS &+ & + &+&- &- &-  &  MR$_3$-CISDT     &  6.1$^d$    &  -189$^d$      &       &  &   \\   
 10 &  18          &3 OS &+ & + &+&- &+,- &-  &  MR(3)-CISDT           &     &       & 13.3   & -104  & -0.346   \\    
 \hline\hline
 6a  &  18            & Single ref.      &+&+&-&-&-&-& CISD                   &     &       &  12.4 & -48 & -0.329 \\    
 6b  &  18            & Single ref.      &+&+&+&-&+,-&-& CISDT                  &     &       &  14.5 & -117& -0.342 \\    
 6   &  18            & Single ref.      &+ & + &+&+ &+ &+  & CISDTQ                 &     &       &  12.1 & -83 & -0.381 \\    
 \hline
 11  &  18            & Single ref.      &+&+&(+)&(+)&(+)&(+)& CCSD                     &     &        & 12.8  & -115& -0.368 \\ 
 12 &  18            & Single ref.      &+&+&+&(+)&+,(+)&(+)& CCSD(T)                  &     &        & 11.0  & -103& -0.387 \\ 
 13 &  18            & Single ref.      &+&+&+&(+)&+,(+)&(+)& CCSDT                    &     &        & 10.6  & -96 & -0.387 \\  
 14 &  18            & Single ref.       &+ & + &+&+ &+ &+  & CCSDTQ                   &     &        & 10.4  & -93 & -0.388 \\   
\hline 
 15 &  18            & 3 OS+all virt. &+ & + &+&+ &(+) &(+)  & MR(${\infty}$)-CCSD       &     &        & 10.0  & -98 & -0.374 \\  
 16 &  18            & 3 OS+all virt. &+ & + &+&+ &+ &(+)  & (MR(${\infty}$)-CCSDT)$_4$&     &        & 10.3  & -91 & -0.388 \\  

\hline\hline
 
\end{tabular}

~\\

\flushleft
$^a$ The absolute values of the considered parameters are given in brackets for the two-electron CISD method only.

$^b$ The difference $\sim11$~GV/cm with the corresponding value of Fleig and Nayak is
mainly due to the spin-orbit interaction, see section ``Results and discussions''.

$^c$ Values of  \Eeff\ and A$_{||}$ calculated within this method are considered as final in Ref. \cite{Fleig:14}.

$^d$ vDZ basis set was used in Ref.~\cite{Fleig:14} for this calculation. 
The given values for MR$_3$-CISDT obtained as 18e-MR$_3$-CISDT/vDZ $-$ 18e-MR$_3$-CISD/vDZ $+$ 18e-MR$_3$-CISD/vTZ $-$ 2e-CISD/vTZ.
Values for MR$_3^{+T}$-CISD obtained in the similar way.
The other results from Ref.~\cite{Fleig:14} presented here are obtained with the vTZ basis set. 
% Therefore, comparison with other results is not clear, but given for completeness.

$^e$ The virtual space of spinors was limited by the 5 Hartree cutoff. 

%$^f$ Due to a different technology of correlation treatment it was not possible to perform the MR$_3^{+T}$-CISD calculation.

$^f$ MR$_3^{+T}$-CISD includes all excitations of the MR(3)-CISD plus triple excitations of the closed-shell valence electrons to the open shell space (not to the virtual space).

$^g$ MR$_3^{+TQ}$-CISD includes all excitations of the MR(3)-CISD plus triple and quadruple excitations of the closed-shell valence electrons to the open shell space (not to the virtual space).

\end{table*}

Table \ref{TCorr} presents the results of correlation calculations with different methods. The collected values of \Eeff, A$_{||}$ and total electron energy are given with respect to the corresponding values obtained within the two-electron configuration interaction approach with single and double excitations (2e-CISD) calculation for convenience (e.g., the spin-orbit contribution to \Eeff\ is about 11 GV/cm, see below), however, the absolute values of the considered parameters are given in brackets for the reference 2e-CISD method (see first line in table \ref{TCorr}). In the 2e-CISD calculation the closed-shell orbitals are excluded from correlation treatment. The results of lines 2--6 are given in the order of improving the level of accounting for the electron correlation. In particular, the MR(12)-CISD method includes all excitations of the MR(3)-CISD method, while the MR(${\infty}$)-CISD method includes all excitations of the MR(12)-CISD method. The methods differ by the number of orbitals included into the active space of MR-CI method. In paper \cite{Fleig:14} 24 active spinors (12 Kramers pairs) are taken at most. The set of spinors includes six spinors (three Kramers pairs) from the open-shell space defined above as well as 18 spinors from the virtual spinor space. The detailed composition of the spinors is given in Ref.~\cite{Fleig:14}. In our scalar-relativistic calculations it was not possible to choose the same orbitals due to their different ordering and composition within the one-component approach. Therefore, we have included all the virtual orbitals into the active space, i.e., we used the MR(${\infty}$)-CISD method. It takes account of the following excitations:
 (i) one- and two-fold excitations from the open-shell space to all the virtual states, which can be denoted as 1v-, 2v- excitations, respectively;
 (ii) one- and two-fold excitations of 16 closed-shell valence electrons to all the open-shell and virtual orbitals, which are denoted as 1o- and 2o- excitations, respectively;
(iii) simultaneous ``connected''  diagrams combining different types of excitations, i.e., 2o2v, 2o1v, 1o2v and 1o1v.
In line~5 the results of (MR(${\infty}$)-CISDT)$_4$ calculation are given
%.
   (it is the MR(${\infty}$)-CISDT method limited by 4-fold excitations with respect to the leading reference determinant).
It includes all the excitations treated by MR(${\infty}$)-CISD  and, besides, the following two types of excitations: 
(i) 3-fold excitations from the valence
to open-shell and virtual spaces, 3o-excitations;
(ii) simultaneous combinations of 3-fold excitations from closed-shell valence space and 
1-fold excitation from the open shells (open-shell orbitals, occupied in the leading reference determinant), i.e., 3o1v. 
Note about the difference between designations in the present paper and those in Ref.~\cite{Fleig:14}. In particular, MR(3)-CISD and MR$_3$-CISD mean the same. However, the MR(3)-CISDT method includes 1o2v, 2o1v, 3o excitations types in addition to those in the MR(3)-CISD method, i.e., all the types of triple excitations, while MR$_3$-CISDT includes only 1o2v, 2o1v types of triple excitations.
No triple excitations from the closed-shell valence space to virtual orbital are taken into account in the MR$_3$-CISDT method. For excitation types of these and other considered methods see table \ref{TCorr}.
Finally, line 6 gives the results of CISDTQ calculations. The method includes all excitations of the (MR(${\infty}$)-CISDT)$_4$ method as well as 4-fold excitations from the closed-shell valence space, 4o.

In Ref.~\cite{Fleig:14} the importance of triple excitations from the closed-shell valence space to six open-shell spinors (three Kramers pairs) was studied. For this, the results of the MR$_3^{+T}$-CISD calculation was compared to the MR(3)-CISD calculation. The MR$_3^{+T}$-CISD method includes all excitations of the MR(3)-CISD plus triple excitation of the closed-shell valence electrons to the open shell space (no 3-fold excitations of the closed-shell valence electrons to virtual space, 3o, are allowed). 
The contribution was small (see lines 7 and 2). Due to a different technology of correlation treatment (within the single-active space MR-CI rather than the generalized active space CI used in Ref.~\cite{Fleig:14}) we were not able to perform exactly the same type of calculation, but we performed the MR$_3^{+TQ}$-CISD calculation which also includes quadruple excitation from closed valence to open-shell space. In this case we have found that the triple and quadruple excitations are also not very important (compare lines 8 and 2). 
``Individual'' contribution of 3-fold excitations from the closed-shell valence space to open-shell and \textit{virtual} spaces, 3o-excitations, can be roughly estimated from comparison of lines 10 and 9 (unfortunately, the results are obtain within different technologies/basis sets and should be compared with caution, though, for the MR(3)-CISD method the approaches give very close results, see line 2).
A ``formal difference'' between MR(3)-CISDT (line 10, this paper) and MR$_3$-CISDT (line 9, Ref.~\cite{Fleig:14}) gives contribution of 3o-excitation to \Eeff\ of about +7.2 GV/cm. Thus, the excitations can almost completely compensate big contribution of 1o2v, 2o1v excitations found in Ref.~\cite{Fleig:14} from comparison of MR$_3$-CISDT and MR$_3$-CISD (see lines 2,9).
Comparison of the (MR(${\infty}$)-CISDT)$_4$ and MR(${\infty}$)-CISD results (lines 5 and 4) shows that  
%   the triple excitations from the closed valence space to open-shell and \textit{virtual} spaces are rather important for \Eeff\ and should be taken into account.
the sum of 3o- and 3o1v- excitations
%   (note that both these methods also account for the 1o2v, 2o1v, 2o2v excitations with respect to the MR(3)-CISD method)
also significantly contribute to \Eeff\ (probably, with different signs). 
Note that both these methods take also into account the 1o2v, 2o1v, 2o2v excitations with respect to the MR(3)-CISD method.
 
Comparison of lines 6 and 5 gives contribution of 4o excitation type which further increases \Eeff\ with respect to MR(${\infty}$)-CISD.
Thus, from lines 4 and 6 it is clear that the MR(${\infty}$)-CISD method is insufficient to treat electron correlation in ThO accurately. 
%   The same is true also for the MR(12)-CISD method which was considered in Ref.~\cite{Fleig:14} as the final (most elaborate) method of correlation treatment. 
%
This can be also seen from comparison of total electronic energies. The difference of energies within the MR(${\infty}$)-CISD and MR$_{3}$-CISD methods is 0.006 Hartree, while E(CISDTQ)$-$E(MR(${\infty}$)-CISD) is 0.046 Hartree.

Due to its exponential ansatz the single-reference coupled-clusters approaches CCn (n means that at most n-fold cluster amplitudes are included in the correlation treatment) can effectively treat higher-order excitations than the corresponding single-reference CIn method, e.g., the CCSD method effectively accounts for very important disconnected four-fold \cite{Bartlett:95} and higher-order excitations. In most cases the single-reference CC approaches work well when there are no large 1-fold excitations  (so called ``t1-diagnostic'') \cite{Bartlett:95}. The maximal 1-fold amplitude for the 38-electron 1c-CCSD calculation with Lbas basis set was found to be 0.05 that is small enough. Large amplitudes can lead to bad convergence in series of the CCn calculations. From table \ref{TCorr} it follows that in the present case single-reference CC series converges well, see lines 11-14.  In particular, the parameters calculated at the CCSDTQ level (which explicitly includes all the excitations of the CISDTQ method) almost negligibly differ from the parameters calculated at the CCSDT level. Note that the CC energy is not variational and it formally can not be treated as a measure of accuracy of the results. However, in the case of inclusion of all possible cluster amplitudes it will converge to the full-CI energy. The almost coincidence of the CCSDT and CCSDTQ energies also favours the CC convergence. In contrast, the convergence of energy (and other properties) in the single-reference CIn series, CISD-CISDT-CISDTQ (lines 6a, 6b, 6) is rather poor and is not achieved.

To test the importance/unimportance of accounting for multireference effects within the coupled-clusters methods we have also performed multireference coupled-clusters calculation using state-selective ansatz \cite{Kallay:2}. The results of the MR(${\infty}$)-CCSD and (MR(${\infty}$)-CCSDT)$_4$ (by analogy with the MR(${\infty}$)-CISD and (MR(${\infty}$)-CISDT)$_4$ methods, correspondingly) are presented in lines 15, 16 in table \ref{TCorr}. Comparing results of configuration interaction calculations within the CISD, MR(${\infty}$)-CISD, (MR(${\infty}$)-CISDT)$_4$, CISDTQ methods (lines 6a, 4, 5, 6) and corresponding coupled-clusters results within the CCSD, MR(${\infty}$)-CCSD, (MR(${\infty}$)-CCSDT)$_4$, CCSDTQ methods (lines 11,15,16,14) one can see that CC-series is much more stable. Actually, inclusion of triple cluster amplitudes within the single-reference CCSDT or CCSD(T) leads to a result which is very close to that of the most elaborate considered CC method (CCSDTQ) in contrast to CISDT vs CISDTQ.

Finally, one can interpret the results in the following way: high-order (three- and four-fold) excitations from the closed and open valence spaces to the space of virtual orbitals with respect to the leading reference determinant are important. Some part of the excitations can be considered within the MR(${\infty}$)-CISD method (due to the multireference nature it includes some types of three- and four-fold excitations with respect to the leading reference determinant (1o2v, 2o1v, 2o2v-excitations, see table \ref{TCorr}). However, it is not enough because important three- and four-fold excitations from the closed-shell valence space (3o,3o1v,4o) are missing in the method. On the other hand effective inclusion of higher-order excitations (formally, up to an infinite order, but for real systems the order is limited by the number of correlated electrons and the number of iterations in the coupled-clusters calculation) within the CCSD(T) approach is  ``almost sufficient'' to obtain converged results. 
%   This method includes three-fold excitations from closed and open-shell space as well as four- and upper- fold disconnected excitations. On the basis of the above analysis we conclude that the theoretical uncertainty due to limited level of accounting for electron correlation of the CCSD(T) method is smaller than 1\% while theoretical uncertainty of the MR(${\infty}$)-CISD

Theoretical uncertainties of \Eeff\ and other considered parameters can
also
 come from the specific choice of one-electron spinors which are used to construct Slater determinants while the total space of the spinors is not changed.
There are no clear criteria to choose the ``best'' spinors. One can use spinors produced in the Hartree-Fock method, complete active space self-consistent field method, natural spinors produced by diagonalizing the one-electron density matrix obtained at some preliminary (previous iteration, etc.) calculation. Obviously, the smaller dependence of the results of correlation calculation on the choice of the spinor set the better. Here we test the stability of the single-reference CCSD and CCSD(T) methods as well as the multireference MR(3)-CISD method due to small change of one-electron spinors.
Note that in the calculations below we use different sets of one-electron spinors. However, in all of the calculations as well as in our previous calculations in Ref.~\cite{Skripnikov:13c} the leading reference determinant (``Fermi vacuum'')  corresponds to the $7s^1 6d_{\delta}^1$(Th) configuration and not to the $7s^2$ one as was mistakenly stated in Ref.~\cite{Fleig:14} with respect to our previous calculations \cite{Skripnikov:13c}. We have performed two-component (i.e., including the spin-orbit effects) 18-electron calculations with the Mbas basis set using the following two choices of sets of one-electron spinors: 
(i) spinors obtained in the two-component average-of-configuration Hartree-Fock calculation for two electrons in the six spinors (three Kramers pairs) which corresponds to 7s, 6d$_\delta$ of Th with all other electrons restricted to closed shells (i.e., the same scheme that was used in Ref.~\cite{Fleig:14});
(ii) spinors produced in the two-component Hartree-Fock calculation of the ground $^1\Sigma^+$ state of ThO.
Due to the absence of one-to-one correspondence between spinors from the two sets it is hard to estimate the uncertainty of results of the multireference methods. Therefore we have not performed MRCI calculations with larger active space than that in MR(3)-CI ones. The results are given in table \ref{RefereceChoice}.

\begin{table}[!h]
\caption{
The calculated uncertainties of \Eeff, A$_{||}$ and $W_{T,P}$ due to different choices of set of one-electron functions (spinors)
within the two-component (2c) approaches using MBas basis set.
The first set is obtained in the two-component average-of-configuration Hartree-Fock calculation for two electrons on six spinors (three Kramers pairs) which corresponds to 7s, 6d$_\delta$ of Th with all the other electrons restricted to closed shells. The second set is generated in the two-component Hartree-Fock calculation of the $^1\Sigma^+$ state of ThO.
}
\label{RefereceChoice}
\begin{tabular}{ l  c  c  c  c  }
\hline\hline
  Method         & $d$,  & \Eeff, & $W_{T,P}$, & A$_{||}$,  \\
                 & Debye & GV/cm  & kHz        & $\frac{\mu_{\rm Th}}{\mu_{\rm N}}\cdot$MHz \\
\hline 
18e-2c-MR(3)-CISD &  0.20 (5\%)   &  -5.8 (7\%) & -11  (10\%)      & -306 (11\%)  \\    
18e-2c-CCSD     & -0.01 (0\%)  &  -0.3 (0\%)  & -1   (0\%)      & -17  (1\%) \\    
18e-2c-CCSD(T)     & +0.01 (0\%)  &  -0.2 (0\%)  & -1   (0\%)      & +2  (0\%) \\    

\hline\hline
\end{tabular}
\end{table} 

It follows from table \ref{RefereceChoice} that the single-reference coupled-clusters calculations are almost irrelevant to the choice of a spinor set (see also below). Such a small dependence of the results is due to a well-known feature of the coupled-clusters approaches which   include single clusters amplitudes (CCS, CCSD, CCSD(T) etc.). The latter are required to account for relaxation effects (e.g., see \cite{Bartlett:95}). However, the values of the considered parameters within the multireference MR(3)-CISD approach have strong dependence on the choice of one-electron spinors. In particular, the uncertainty of MR(3)-CISD calculation is 7\% for \Eeff\ and 11\% for A$_{||}$.

Due to aforementioned lacks of the considered MRCI approaches (strong dependence on the choice of set of one-electron spinors, big theoretical uncertainties with respect to higher-order correlation methods) we chose the coupled-clusters approaches as the basic ones for the two-component calculations given below like that in our previous paper \cite{Skripnikov:13c}.

% Finally, it should be noted that there is one more possible source of uncertainty of the AiC properties calculated in the present paper caused by approximate nature of the non-variation restoration procedure the impossibility to use full version of the generalized effective core potential due to limitations of the used codes.
%According to our estimations based on different choices of the equivalent basis sets it can achive 5\% for \Eeff.
%Note that the value of \Eeff\ obtained within two-component 2-electron CISD and 18 electron MR3-CISD methods using Mbas are bigger that the corresponding values obtained by Fleig and Nayak in Ref. \cite{Fleig:14} by 2.7 GV/cm (3\%),
%while hyperfine structure constant is lower by absolute value 120~$\frac{\mu_{\rm Th}}{\mu_{\rm N}}\cdot$MHz (4\%).
%One cannot exactly explain the difference due to different basis sets, possible contribution of the Breit interaction considered in this paper, etc. However, we include it in the final theoretical uncertainty of the Aic properties.

%---------------------------------
\section{Outer-core contributions}

Due to the size-extensive property of the used coupled-clusters approaches it is possible to investigate contribution to \Eeff\ and other parameters under consideration from different shells of Th. This feature was used to  evaluate the contribution of the outer core electrons of Th (5s-5d shells) and 1s(O). For this we have performed the 2c-CCSD(T) calculations within MBas basis set where these electrons were frozen (i.e., only 18 electrons were correlated) and compared the obtained value of \Eeff\ with that from 38e-calculation. The results are given in table \ref{TFrozen}.

\begin{table}[!h]
\caption{
The calculated values of the $H^3\Delta_1\to X^1\Sigma^+$ transition energy ($T_e$), molecule-frame dipole moment ($d$), effective electric field (\Eeff), parameter of the \PSen\ interaction ($W_{T,P}$) and hyperfine structure constant (A$_{||}$) for the $H^3\Delta_1$ state of ThO using the 2c-CCSD(T) method and Mbas basis set.
}
\label{TFrozen}
\begin{tabular}{ l  c  c  c  c  c  c}
\hline\hline
\# of act.  &  Virt.\ orb.      & $T_e,$     & $d$,  & \Eeff, & $W_{T,P}$, & A$_{||}$,  \\
 electrons  & cutoff & \cm  & Debye & GV/cm & kHz & $\frac{\mu_{\rm Th}}{\mu_{\rm N}}\cdot$MHz \\
\hline
%  18 & $\infty$           & {} 4526        & {} 4.16    & {} 78.9  & 108  & -2894  \\  
%  38 & 5 a.u.             & {} 4859        & {} 4.20    & {} 79.6  & 109  & -2879 \\  
%  38 & $\infty$           & {} 4994        & {} 4.27    & {} 82.8  & 114  & -3004 \\    
  18 & $\infty$           & {}  4983      & {}  4.13   & {} 77.4  & 107  & -2844  \\  
  38 & 5 Hartree          & {}  5367      & {}  4.18   & {} 78.4  & 108  & -2814 \\  
  38 & $\infty$           & {}  5525      & {}  4.21   & {} 81.7  & 112  & -2922  \\

\hline\hline
\end{tabular}
\end{table}

From table \ref{TFrozen} one can see that the correlation contribution to \Eeff\ from the outer core electrons calculated as a difference between the 38- and 18-electron calculations, (see lines 1 and 3) is 4.3~GV/cm. In these calculations no restriction on the active space of virtual spinors was used. Fleig and Nayak have performed comparative 36-electron and 18-electron MRCI calculations using the MR(3)-CISD method \cite{Fleig:14} and, as a result, they have concluded that the outer-core electrons negligibly contribute to \Eeff. However, in those calculations the active space of virtual spinors was limited by orbital energy of 5~Hartree
%    It was earlier shown on different molecules \cite{Titov:06amin} that the outer core relaxation/correlation effects can contribute to the T,P-odd effects considerably, up to about 50\% from contribution of the valence electrons.
%
Such a truncation seems us rather dangerous since the orbital energies of outer core $5s$ and $5p_{1/2}, 5p_{3/2}$ states are about $-12, -9$ and $-8$~a.u., correspondingly, whereas the energies of $5d_{3/2}, 5d_{5/2}$ states are about $-4$~Hartree. To take account of their correlation/relaxation effects reliably one should involve the virtual states which are largely localized at the same space region, i.e., their orbital energies (with positive signs) would be at least about as twice as large by absolute value as those of corresponding outer core states.
To investigate the consequences of  the 5~Hartree cutoff we have performed additional calculation with the restriction of active space of virtual spinors, see line 2 of table \ref{TFrozen}. It follows from the table that the limitation of active space leads to the similar consequence as freezing outer-core electrons, i.e., \Eeff\ decreases by 3.3 GV/cm. Analogues behavior can be found for A$_{||}$ and $W_{T,P}$. Basing on this study one can suggest that all 18-electron calculations of \Eeff\ have theoretical uncertainly (underestimation) more than 5\%.

Within the MR(3)-CISD calculations performed in Ref.~\cite{Fleig:14} using vDZ basis set the difference between transition energies obtained within the 36-electron calculation ($1s^2$(O) is frozen) with virtual space restricted to 5~Hartree and 18-electron calculation with virtual space restricted to 38~Hartree was found to be -90~\cm. Our corresponding 36-electron and 18-electron two-component MR(3)-CISD calculations with the Mbas basis set gave the value of the difference equal to -145 \cm\ (the results do not coincide exactly due to different basis sets, possible contribution of Breit interaction considered in the present work, etc.). We think that the contribution of the correlation of outer-core electrons can not be considered in such a way because the MR(3)-CISD method is nor size-extensive nor size-consistent \cite{Bartlett:95}.
%   (ii) the divergence of the transition energy from the experimental value is large -- about 30\%, see  Ref. \cite{Fleig:14}.
Our size-extensive and size-consistent coupled-clusters calculations showed that the outer-core electrons contribute about $+540$~\cm\ (10\%) into the transition energy for the case of non-restricted virtual space (compare lines 1 and 3 from table \ref{TFrozen}). It should be noted that the contribution to the excitation energy is rather stable with respect to the coupled-clusters method. Scalar-relativistic calculations with the CBas basis set showed that the contributions calculated at the CCSD(T), CCSDT and CCSDT(Q) methods coincide within 60~\cm.

%*** Meyer =104, Fleig = 71.7-75.2 (74.1)
% , however, it is more than 10\% larger than the values given in Ref.~\cite{Fleig:13Aa}. 
% Method    Eeff
% UHF        86
% ROHF       58
% U-CISD     78
% RO-CISD    66 
% U-CCSD     72
% RO-CCSD    72
%
%*****

%--------------------------------
\section{Results and discussions}

On the basis of the study given above we have chosen the coupled-clusters approach as the method to account for the electron correlation. Due to considerable contribution to \Eeff\ (and other considered parameters) from the outer-core electrons, the outermost 38 electrons of ThO were correlated. Our final calculation scheme for the transition energy between the ground $X ^1\Sigma^+$ and excited $H^3\Delta_1$ states as well as the molecule-frame dipole moment, g-factor and AiC parameters (\Eeff, $W_{T,P}$ and $A_{||}$) for the $^3\Delta_1$ state of ThO included the following steps:
(i) calculation of the main contributions within the two-component CCSD(T) method
and using the MBas basis set;
(ii) calculation of basis set correction within the scalar-relativistic CCSD(T) method;
(iii) calculation of the correction on iterative triple and non-iterative quadruple cluster amplitudes;
(iv) analysis of different contributions and error estimates.

To calculate the correction on the basis set enlargement we have performed {\it (i)} 
38-electron
scalar-relativistic CCSD(T) calculation using the same basis set as was used for the two-component calculation, MBas, and {\it (ii)} 
38-electron
scalar-relativistic CCSD(T) calculation using the extended basis set Lbas. The corrections are estimated as differences between the values of the corresponding parameters. Note that {\it no cuts} of the active space of orbitals by energy were done in the correlation calculations. In particular, in the 38-electron CCSD(T) calculation with the LBas all 1204 spin-orbitals were considered explicitly.

To compute the correction on inclusion of iterative triple and non-iterative quadruple cluster amplitudes we have performed: 
(i) 18-electron two-component calculation within the CCSD(T) method and (ii) 18-electron two-component calculation using the coupled-clusters method with single, double, triple and non-iterative quadruple cluster amplitudes, CCSDT(Q). The CBasSO basis set was used in these calculations. The final value of a considered parameter $Y$, where Y=$T_e$, $d$, $G_{\parallel}$, \Eeff, $W_{T,P}$, A$_{||}$  of the $H ^3\Delta_1$ state of ThO was obtained as:
\begin{equation}
\begin{array}{l}
Y({\rm FINAL}) = Y(\mbox{38e-2c-CCSD(T), MBas}) + \\
                 Y(\mbox{38e-1c-CCSD(T), LBas}) - Y(\mbox{38e-1c-CCSD(T), MBas}) \\
                 + Y(\mbox{18e-2c-CCSDT(Q), CBasSO}) \\
                  - Y(\mbox{18e-2c-CCSD(T), CBasSO})
\end{array}
\end{equation}
Corrections on the basis set and correlation effects are given in table \ref{TResults}.
% In the scalar-relativistic CCSD(T) calculations we have found that \Eeff\ only very slightly depends on the internuclear distance: it has a maximum at 3.48 a.u., while at the interatomic distances of 3.4 a.u.\ and 3.56 it decreases by 0.1 GV/cm with respect to the maximal value. 
%  
\begin{table*}[!h]
\caption{
The calculated values of the $H^3\Delta_1\to X^1\Sigma^+$ transition energy, molecule-frame dipole moment ($d$), effective electric field (\Eeff), parameter of the \PSen\ interaction ($W_{T,P}$), hyperfine structure constant (A$_{||}$) and g-factor (G$_{||}$) of the $H ^3\Delta_1$ state of ThO using different methods.
In four- (4c-) and two- (2c-) component methods spin-orbit effects are taken into account while in one- (1c-) component methods they are excluded for valence electrons.
%  The GRECP calculations were performed with (1c) and without (2c) taking into account the spin-orbit effects.
}
\label{TResults}
\begin{tabular}{l l  r  r  r  r  r  c}
\hline\hline
Ref & Method & $T_e,$     & $d$,  & \Eeff, & $W_{T,P}$, & A$_{||}$,   & $G_{\parallel}$  \\
    &       & $cm^{-1}$  & Debye & GV/cm & kHz & $\frac{\mu_{\rm Th}}{\mu_{\rm N}}\cdot$MHz & \\
%\hline       
%        
\hline
%  2c-CCSD                 & {} 4994 & {} 4.27    & {} 82.8  & 114  & -3004 \\  
%  2c-CCSD(T)              & {} 5822 & {} 4.23    & {} 81.7  & 112  & -2927 \\ 
%\hline  
%  basis set         & {} -208  & {} 0.09   & {} -0.33 & 0  & 29    \\ 
%  ~~correction        & {}       & {}        & {}       &    &       \\ 
%  correlation       & {}  1    & {}  --    & {} +0.32 & -- &  --   \\  
%  ~~correction        & {}       & {}        & {}       &    &       \\   
%  FINAL                   & {} 5615 & {}  4.14   & {} 81.6  & 112 & 2956 \\  

\hline
Ref. \cite{Fleig:14} &  2e-4c-CISD/vTZ$^{a,b}$   & {} 5929
$^{c}$
 & {}  ---   & {} 68.5  & ---     &  -2809  & ---   \\  
%\\
%                     &  18e-4c-MR(3)-CISD/vDZ$^{a,b}$   & {} 4535 & {}  ---   & {} 80.8  & ---     &  -2851 & ---    \\  
%                     &  18e-4c-MR(3)-CISD/vTZ$^{a,b}$   & {} 3832 & {}  ---   & {} 81.0  & ---     &  -2871 & ---    \\  
%                     &  18e-4c-MR(3)-CISD/vQZ$^{a,c}$   & {} 3643 & {}  ---   & {} 80.7  & ---     &  -2884 & ---    \\                  
%                          
                     &  18e-4c-MR(9)-CISD/vDZ$^{a,b}$   & {} 5703 & {}  ---   & {} 73.4  & ---     &  -2936 & ---    \\                       
                     &  18e-4c-MR(9)-CISD/vTZ$^{a,e}$   & {} 5125 & {}  ---   & {} 76.3  & ---     &  -2947 & ---    \\                                                                                                         
%\\
                     &  18e-4c-MR(12)-CISD/vTZ$^{a,b}$   & {} 5410 & {}  ---   & {} 75.2  & ---     &  -2976 & ---    \\                                         
\hline
%This work &  18e-2c-MR(3)-CISD$^a$   & {} 3647 & {}  3.90   & {} 83.7  & 115     &  -2749     \\  
This work &  2e-2c-CISD$^a$   & {}  5650 $^a$ & {}  4.50   & {} 70.6  & 97     &  -2693 & ---    \\  
\hline
Ref. \cite{Skripnikov:13c} &  38e-2c-CCSD(T)   & {} 5808 & {}  4.32   & {} 84  & 116     &  -2885 & ---    \\  
   &+basis set correction 
$^e$
   &  & & &  & &   \\  
\hline
This work  & 38e-1c-CCSD       & {} 6321 & {}  4.30   & {} 72.9  & 102     & -3040    & ---  \\
           & 38e-1c-CCSD(T)    & {} 6698 & {}  4.23   & {} 71.0  & 100  & -2961   & ---   \\
\\           
           &  38e-2c-CCSD      & {} 5210 & {}  4.28   & {} 83.2  & 114     &  -2998 & 0.003~    \\  
&  38e-2c-CCSD(T)              & {} 5525 & {}  4.21   & {} 81.7  & 112     &  -2920 & 0.006~    \\ 
%\hline  
&  basis set correction              & {} -208 & {} -0.09   & {} -0.3  & 0  & -29  & ---   \\ 
%&  ~~correction            & {}      & {}         & {}       &    &    &    \\ 
&  correlation correction    & {}  +86 & {} +0.08   & {} +0.1  & 0  & +1 & 0.001~  \\  
%&  ~~correction            & {}      & {}         & {}       &    &   &    \\   
&  \textbf{FINAL}                   & {} 5403 & {}  4.19   & {} 81.5  & 112& -2950 & 0.007~  \\

\hline  
  Experiment &  & {} 5321 & {} 4.24{$\pm$}0.1 & {} ---& {} --- & --- &  0.0083  \\ 
   & & {}  \cite{Huber:79,Edvinsson:84} & {} \cite{Vutha:2011} & {} & {}  & & \cite{Petrov:14,Kirilov:13}  \\  
 
\hline\hline
\end{tabular}
\\
\flushleft
$^a$ 
To compare with \cite{Fleig:14} we set R(Th-O)=3.48 a.u.\ for both $^1\Sigma$ and $^3\Delta_1$ states;
the vertical excitation energy was corrected by -100~\cm\
(``non-parallelity correction''),  active space of virtual orbitals was limited to 38~Hartree.
For other reported calculation we used experimental equilibrium distances and no cutoff for virtual orbitals.

$^b$ vDZ basis set, see \cite{Fleig:14}.

%$^c$ vTZ basis set, see \cite{Fleig:14}.
%
%$^d$ vQZ basis set, see \cite{Fleig:14}.
%
$^c$ Note that at the 18e-4c-MR(3)-CISD level $T_e$ value calculated within vQZ basis set is lower by 189 \cm\ than the value calculated within vTZ basis set, see \cite{Fleig:14}.
No Breit interaction is accounted for in the calculation.
Calculations reported in this work effectively account for main part of Breit interaction within the GRECP operator.

$^d$ vTZ basis set, see \cite{Fleig:14}.

$^e$ In Ref.~\cite{Skripnikov:13c} smaller basis sets were used both for two-component (to get the leading  contributions) and one-component (to get basis set corrections) calculations than in the present paper, see text.

\end{table*} 
It follows from the table that the calculated values of transition energy and molecule-frame dipole moment are in a very good agreement with the experimental data from Refs.~\cite{Huber:79, Edvinsson:84}.
%, the  deviation is on the level of accuracy early attained by our group for compounds of transition metals and lanthanides.
%It is also in a good agreement with previous calculation performed in Ref.~\cite{Paulovic:03}.
The evaluated g-factor is in a reasonable agreement with experimental data \cite{Petrov:14,Kirilov:13} though, basing on its convergence with increasing the level of electron 
correlation we estimate our theoretical uncertainty for g-factor as 20\%.
Rather good agreement for  $T_e,$ and A$_{||}$  between 18e-4c-MR(12)-CISD/vTZ from Ref.~\cite{Fleig:14}
and the present one (38e-2c-CCSD(T)+T(Q) + basis set correction) should be considered as a ``random coincidence'' basing on the consideration given in the above sections.
% Note also, that in general A$_{||}$ is more stable due to 
%(uncertainty or MRCI due to reference choice, absence of important 3o, 3o1v, 4o and other higher-order extitations, absence of outer-core contribution)
%
For all the considered AiC properties, the corrections on the basis set enlargement and increasing the level of accounting for correlation effects (see above) are rather small ($<$1\%). In particular, a correction to \Eeff\  on inclusion of iterative triple and non-iterative quadruple cluster amplitudes is +0.1 GV/cm, i.e., it is even smaller than correction on inclusion of iterative triple and quadruple cluster amplitudes in the scalar-relativistic case, see table \ref{TCorr}.
\footnote{Due to computational/program limitations it was not possible to perform CCSDTQ-calculation in the two-component relativistic case}.

%The \Eeff\ value calculated within the scalar-relativistic 1c-CCSD(T) with the MBas basis set is 70.8 GV/cm. Comparing it to the two-component value from table \ref{TResults} one obtains the contribution from the spin-orbit interaction of outer-core and valence electrons of about 10.9 GV/cm. 
%For ThO we have found rather big contribution of the spin-orbit effects to \Eeff, about 11~GV/cm (compare 1c-CCSD(T) and 2c-CCSD(T) results in table \ref{TResults}). For the hyperfine structure constant its contribution is not so dramatic.
To test an importance of accounting for spin-orbit effects in ThO for AiC properties, table \ref{TResults} presents the results of one-component (scalar-relativistic) 38e-CCSD and 38e-CCSD(T) calculations.
We have found rather big contribution of the spin-orbit effects to \Eeff, about 10-11~GV/cm (compare 1c-CCSD and 2c-CCSD or 1c-CCSD(T) and 2c-CCSD(T)  results). For the hyperfine structure constant its contribution is not so dramatic.

In the present paper we have used larger basis sets both for the two-component calculation (MBas) and one-component calculations (Lbas) with respect to the corresponding basis sets that were used in our previous work \cite{Skripnikov:13c}. One should note that if we take the values obtained in \cite{Skripnikov:13c} within the 2c-CCSD(T) approximation and apply a correction on the new large Lbas basis set we get \Eeff=81.5 that is quite close to the value obtained here (without correlation correction).

Basing on comparison of calculated \Eeff\ values within the 18e-4c-MR(3)-CISD method using the vDZ, vTZ and vQZ basis sets (80.8 GV/cm, 81.0 GV/cm and 80.7 GV/cm, respectively) it was concluded in Ref.~\cite{Fleig:14} that the vTZ basis set is sufficient to attain a good value for \Eeff. One should note, however, that within a more accurate 18e-4c-MR(9)-CISD method the values of \Eeff\ differ by about 2.9 GV/cm for the vTZ vs.\ vDZ basis sets used
(see table \ref{TResults}) that is not negligible (there are no vQZ results reported for MR(9)-CISD).
%   The difference could be even more significant for the case of higher-order excitations taken into consideration.
For a completeness, to make a certain ``direct'' comparison with the results of Ref.~\cite{Fleig:14} we chose the method which allows one to ``minimize'' its dependence on the basis set choice when taking the most important correlation effects, the two-electron configuration interaction with single and double excitations (2e-CISD) method. One can see from the table, that both our and Fleig-Nayak results agree reasonably for the transition energy in this case.

Some of possible sources of uncertainties in the AiC properties calculated here are caused by an approximate nature of the non-variation restoration procedure and by impossibility to use the full version of the GRECP operator (because of limitations in the used molecular codes). According to our estimations based on different choices of the equivalent basis sets it can achieve 5\% for \Eeff. Note that the value of \Eeff\ obtained within the two-component 2e-CISD using Mbas basis set are bigger than the corresponding values obtained by Fleig and Nayak in Ref.~\cite{Fleig:14} by 2.1 GV/cm (3\%), while hyperfine structure constant is lower by absolute value 116~$\frac{\mu_{\rm Th}}{\mu_{\rm N}}\cdot$MHz (4\%), see table \ref{TResults}. One cannot exhaustively explain the divergence of the results due to different basis sets, 
%   possible contribution of the Breit interaction considered in this paper
etc. However, we include it to the final theoretical uncertainty of the AiC properties.

Earlier in the text we have mentioned about a near independence of the 18e-CCSD(T) results due to a particular choice of the set of one-electron spinors. For the final results given in table \ref{TResults} for both the $^1\Sigma^+$ and $^3\Delta_1$ states we used the same set of one-electron spinors obtained in the two-component average-of-configuration Hartree-Fock calculation for two electrons in the  six spinors (three Kramers pairs) which corresponds to 7s, 6d$_\delta$ of Th with all the other electrons restricted to closed shells.
%,  i.e. the same scheme that was used in Ref. \cite{Fleig:14}.
To test the stability of the considered parameters due to the choice of the reference determinant we have performed the \textit{38-electron} 2c-CCSD(T) calculations with the MBas basis set using spinors produced in the two-component Hartree-Fock calculation of the $^1\Sigma^+$ state of ThO. The obtained values of $T_e$~(5822 \cm), $d$~(4.23 Debye), $G_{\parallel}$~(0.005), \Eeff~(81.7 GV/cm), $W_{T,P}$~(112 kHz) and $A_{||}$~(-2927  $\frac{\mu_{\rm Th}}{\mu_{\rm N}}\cdot$MHz) almost coincide with the values obtained within the corresponding calculation based on the spinors from the average-of-configuration Hartree-Fock calculation (see table \ref{TResults}) 
as well as in the 18-electron case discussed above.

In the present calculations we have used the Fermi distribution of the nuclear charge model for the Th nucleus. According to our estimations based on comparison of values of the matrix elements of operators (\ref{matrelem}) and (\ref{Apar}) in the basis set of $7s$ and $7p_{\frac{1}{2}}$ functions the uncertainty of \Eeff\ and A$_{||}$ due to the change of the nuclear model of Fermi to the surface distribution of the nuclear charge model is about 1\% and to the point nucleus model is 6\%. Obviously, the latter model is too bad for Th (Z=90). Therefore, we can conclude that the uncertainties caused by the choice of the nuclear charge model in the calculated values of AiC parameters are of order of 1\%.

At the second stage of the used two-step method of calculating the AiC properties a one-center (on Th) four-component one-electron reduced density matrix is obtained.
Using the matrix one can estimate contributions to \Eeff\ from mixing of $s$ and $p$, $p$ and $d$, etc.\ orbitals. Using the desity matrix from 2c-CCSD calculation with the Mbas basis set one concludes that the main contribution to \Eeff\ comes from mixing of $s$ and $p$ orbitals as expected. However, the contribution from mixing of $p$ and $d$ orbitals is equal to +1.5 GV/cm, so, it is not completely negligible. The $d{-}f$ contribution is about -0.3 GV/cm.

%--------------------
\section{Conclusions}

In the present paper a new series of calculations of the transition energy between the ground $^1\Sigma^+$ and the excited $H ^3\Delta_1$ state  ($T_e$), molecule-frame dipole moment ($d$), g-factor ($G_{\parallel}$), effective electric field (\Eeff), parameter of the T,P-odd pseudoscalar$-$scalar electron$-$nucleus interaction ($W_{T,P}$) and hyperfine structure constant (A$_{||}$)  of the $H ^3\Delta_1$ state of the ThO molecule has been performed.
We have compared the calculated parameters with available experimental data and have found a very good agreement.
To our knowledge, the g-factor is calculated for the first time for the system under consideration. In addition a detailed analysis of correlation contributions and uncertainties of the calculated parameters has been performed:
\begin{itemize}
%(i)
\item
 Investigation of the importance of correlation of outer-core electrons within size-extensive methods showed that exclusion of these electrons or setting the restriction on the energy of virtual spinors of 5~Hartree in the 36- or 38-electron calculations leads to underestimation of \Eeff\ by about 4 GV/cm. In particular, it gives the 5\% uncertainty to the 18-electron calculations of \Eeff. The size-extensive correlation contribution of the outer-core electrons to the  $H^3\Delta_1\to X^1\Sigma^+$ transition energy is about +540 \cm\ (10\%).
%(ii)
\item
 An analysis of correlation contributions to the wave-function showed that the results of the 18-electron MR(${\infty}$)-CISD 
%  (and MR(12)-CISD as a consequence) 
can achieve an additional theoretical uncertainty of about 4 GV/cm (5\%) due to absence of triple and quadruple excitations from the closed-shell valence
orbital (spinor) space to the virtual orbital (spinor) space.
 On the other hand the theoretical uncertainty of the CCSD(T) (estimated as a difference between CCSDT(Q) and CCSD(T) results) consideration of AiC parameters is negligible ($<$1\%).
Besides, it was shown that no multireference consideration within the coupled-clusters
approaches is required, i.e., even the single-reference coupled-clusters theory is enough for the system under consideration.
% (iii)
\item
Finally, the uncertainty of the multireference configuration interaction MR(3)-CISD method due to a particular choice of one-electron spinors is 7\% for \Eeff\ and 11\% for hyperfine structure constant. It is negligible for the considered coupled-clusters approaches. 
\end{itemize}

Basing on the above study we conclude that the uncertainty of \Eeff\ equal to 3\%  declared in \cite{Fleig:14} for the final (18-electron MR(12)-CISD) calculation is notably underestimated.

In the new series of 38-electron coupled-clusters calculations  (i.e., the outer-core electrons are correlated)
of the effective electric field and other AiC parameters with respect to our previous study in Ref.~\cite{Skripnikov:13c} we 
(i) have improved level of correlation treatment by estimation the correction on inclusion of iterative triples and non-iterative quadruple cluster amplitudes,
(ii) have considered extended basis set. 
% (iii) estimated importance of multireference treatment within mutlitereference coupled-clusters and showed that it is not important to use the multireference approach for the case of coupled-clusters treatment of ThO in its $^3\Delta_1$ excited state. 
%Study of instability of the used coupled-clusters approach showed that the treatment is almost independent to the choice of orbitals used.
Finally, the obtained \Eeff=81.5 GV/cm and other 
atom-in-compound
properties have more than twice smaller uncertainty than in our previous treatment \cite{Skripnikov:13c}, it is now within 7\%.

The combined scheme of calculation applied here is suggested to be used further to investigate other system prospective to search for electron electric dipole moment such as ThF$^+$ cation \cite{Leanhardt:2011}, etc.

%%%%%%%%%%%%%%%%%%%%%%%%%%%%%%%%%%%%%%%%%%%%%%%%%%%%%%%%%%%%%%%%%%%%%%%%%%%%%%%
%%%%%%%%%%%%%%%%%%%%%%%%%%%%%%%%%%%%%%%%%%%%%%%%%%%%%%%%%%%%%%%%%%%%%%%%%%%%%%%

\section{Acknowledgement}

The reported study was supported by the Russian Science Foundation grant (project No.~14-31-00022).
We are very grateful to our colleagues A.N.Petrov and N.S.Mosyagin for many fruitful discussions on the subject of the present research.

%\bibliographystyle{./bib/apsrev}
%\bibliography{bib/JournAbbr,bib/SkripnikovLib,bib/QCPNPI,bib/TitovLib,bib/Kaldor,bib/PetrovLib,bib/Lomachuk}

\begin{thebibliography}{68}
\expandafter\ifx\csname natexlab\endcsname\relax\def\natexlab#1{#1}\fi
\expandafter\ifx\csname bibnamefont\endcsname\relax
  \def\bibnamefont#1{#1}\fi
\expandafter\ifx\csname bibfnamefont\endcsname\relax
  \def\bibfnamefont#1{#1}\fi
\expandafter\ifx\csname citenamefont\endcsname\relax
  \def\citenamefont#1{#1}\fi
\expandafter\ifx\csname url\endcsname\relax
  \def\url#1{\texttt{#1}}\fi
\expandafter\ifx\csname urlprefix\endcsname\relax\def\urlprefix{URL }\fi
\providecommand{\bibinfo}[2]{#2}
\providecommand{\eprint}[2][]{\url{#2}}

\bibitem[{\citenamefont{Commins}(1998)}]{Commins:98}
\bibinfo{author}{\bibfnamefont{E.~D.} \bibnamefont{Commins}},
  \bibinfo{journal}{Adv.\ At.\ Mol.\ Opt.\ Phys.}
  \textbf{\bibinfo{volume}{40}}, \bibinfo{pages}{1} (\bibinfo{year}{1998}).

\bibitem[{\citenamefont{Ginges and Flambaum}(2004)}]{Ginges:04}
\bibinfo{author}{\bibfnamefont{J.~S.~M.} \bibnamefont{Ginges}}
  \bibnamefont{and} \bibinfo{author}{\bibfnamefont{V.~V.}
  \bibnamefont{Flambaum}}, \bibinfo{journal}{Phys.\ Rep.}
  \textbf{\bibinfo{volume}{397}}, \bibinfo{pages}{63} (\bibinfo{year}{2004}).

\bibitem[{\citenamefont{Khriplovich and Lamoreaux}(2011)}]{Khriplovich:11}
\bibinfo{author}{\bibfnamefont{I.~B.} \bibnamefont{Khriplovich}}
  \bibnamefont{and} \bibinfo{author}{\bibfnamefont{S.~K.}
  \bibnamefont{Lamoreaux}}, \emph{\bibinfo{title}{{CP} Violation without
  Strangeness. The Electric Dipole Moments of Particles, Atoms, and Molecules}}
  (\bibinfo{publisher}{Springer}, \bibinfo{address}{London},
  \bibinfo{year}{2011}).

\bibitem[{\citenamefont{Chupp and Ramsey-Musolf}(2014)}]{Chupp:14gka}
\bibinfo{author}{\bibfnamefont{T.}~\bibnamefont{Chupp}} \bibnamefont{and}
  \bibinfo{author}{\bibfnamefont{M.}~\bibnamefont{Ramsey-Musolf}}
  (\bibinfo{year}{2014}), \eprint{arXiv:1407.1064}.

\bibitem[{\citenamefont{Sandars and Lipworth}(1964)}]{Sandars:64}
\bibinfo{author}{\bibfnamefont{P.~G.~H.} \bibnamefont{Sandars}}
  \bibnamefont{and} \bibinfo{author}{\bibfnamefont{E.}~\bibnamefont{Lipworth}},
  \bibinfo{journal}{Phys.\ Lett.} \textbf{\bibinfo{volume}{13}},
  \bibinfo{pages}{718} (\bibinfo{year}{1964}).

\bibitem[{\citenamefont{Sandars}(1965)}]{Sandars:65}
\bibinfo{author}{\bibfnamefont{P.~G.~H.} \bibnamefont{Sandars}},
  \bibinfo{journal}{Phys.\ Lett.} \textbf{\bibinfo{volume}{14}},
  \bibinfo{pages}{194} (\bibinfo{year}{1965}).

\bibitem[{\citenamefont{Sandars}(1967)}]{Sandars:67}
\bibinfo{author}{\bibfnamefont{P.~G.~H.} \bibnamefont{Sandars}},
  \bibinfo{journal}{Phys.\ Rev.\ Lett.} \textbf{\bibinfo{volume}{19}},
  \bibinfo{pages}{1396} (\bibinfo{year}{1967}).

\bibitem[{\citenamefont{Onischuk}()}]{Onischuk:67}
\bibinfo{author}{\bibfnamefont{V.~A.} \bibnamefont{Onischuk}},
  \bibinfo{note}{preprint JINR, R4-3299 (Dubna, 1967)}.

\bibitem[{\citenamefont{Harrison et~al.}(1969)\citenamefont{Harrison, Sandars,
  and Wright}}]{Harrison:69a}
\bibinfo{author}{\bibfnamefont{G.~E.} \bibnamefont{Harrison}},
  \bibinfo{author}{\bibfnamefont{P.~G.~H.} \bibnamefont{Sandars}},
  \bibnamefont{and} \bibinfo{author}{\bibfnamefont{S.~J.}
  \bibnamefont{Wright}}, \bibinfo{journal}{Phys.\ Rev.\ Lett.}
  \textbf{\bibinfo{volume}{22}}, \bibinfo{pages}{1263} (\bibinfo{year}{1969}).

\bibitem[{\citenamefont{Hinds et~al.}(1976)\citenamefont{Hinds, Loving, and
  Sandars}}]{Hinds:76}
\bibinfo{author}{\bibfnamefont{E.~A.} \bibnamefont{Hinds}},
  \bibinfo{author}{\bibfnamefont{C.~E.} \bibnamefont{Loving}},
  \bibnamefont{and} \bibinfo{author}{\bibfnamefont{P.~G.~H.}
  \bibnamefont{Sandars}}, \bibinfo{journal}{Phys.\ Lett.\ B}
  \textbf{\bibinfo{volume}{62}}, \bibinfo{pages}{97} (\bibinfo{year}{1976}).

\bibitem[{\citenamefont{Hinds and Sandars}(1980)}]{Hinds:80b}
\bibinfo{author}{\bibfnamefont{E.~A.} \bibnamefont{Hinds}} \bibnamefont{and}
  \bibinfo{author}{\bibfnamefont{P.~G.~H.} \bibnamefont{Sandars}},
  \bibinfo{journal}{Phys.\ Rev.\ A} \textbf{\bibinfo{volume}{21}},
  \bibinfo{pages}{480} (\bibinfo{year}{1980}).

\bibitem[{\citenamefont{Labzowsky}(1978)}]{Labzowsky:78}
\bibinfo{author}{\bibfnamefont{L.~N.} \bibnamefont{Labzowsky}},
  \bibinfo{journal}{Sov.\ Phys.--JETP} \textbf{\bibinfo{volume}{48}},
  \bibinfo{pages}{434} (\bibinfo{year}{1978}).

\bibitem[{\citenamefont{Sushkov and Flambaum}(1978)}]{Sushkov:78}
\bibinfo{author}{\bibfnamefont{O.~P.} \bibnamefont{Sushkov}} \bibnamefont{and}
  \bibinfo{author}{\bibfnamefont{V.~V.} \bibnamefont{Flambaum}},
  \bibinfo{journal}{Sov.\ Phys.--JETP} \textbf{\bibinfo{volume}{48}},
  \bibinfo{pages}{608} (\bibinfo{year}{1978}).

\bibitem[{\citenamefont{Gorshkow et~al.}(1979)\citenamefont{Gorshkow,
  Labzovsky, and Moskalyov}}]{Gorshkov:79}
\bibinfo{author}{\bibfnamefont{V.~G.} \bibnamefont{Gorshkow}},
  \bibinfo{author}{\bibfnamefont{L.~N.} \bibnamefont{Labzovsky}},
  \bibnamefont{and} \bibinfo{author}{\bibfnamefont{A.~N.}
  \bibnamefont{Moskalyov}}, \bibinfo{journal}{Sov.\ Phys.--JETP}
  \textbf{\bibinfo{volume}{49}}, \bibinfo{pages}{209} (\bibinfo{year}{1979}).

\bibitem[{\citenamefont{Sushkov et~al.}(1984)\citenamefont{Sushkov, Flambaum,
  and Khriplovich}}]{Sushkov:84}
\bibinfo{author}{\bibfnamefont{O.~P.} \bibnamefont{Sushkov}},
  \bibinfo{author}{\bibfnamefont{V.~V.} \bibnamefont{Flambaum}},
  \bibnamefont{and} \bibinfo{author}{\bibfnamefont{I.~B.}
  \bibnamefont{Khriplovich}}, \bibinfo{journal}{Sov.\ Phys.--JETP}
  \textbf{\bibinfo{volume}{87}}, \bibinfo{pages}{1521} (\bibinfo{year}{1984}).

\bibitem[{\citenamefont{Flambaum and Khriplovich}(1985)}]{Flambaum:85b}
\bibinfo{author}{\bibfnamefont{V.~V.} \bibnamefont{Flambaum}} \bibnamefont{and}
  \bibinfo{author}{\bibfnamefont{I.~B.} \bibnamefont{Khriplovich}},
  \bibinfo{journal}{Phys.\ Lett.\ A} \textbf{\bibinfo{volume}{110}},
  \bibinfo{pages}{121} (\bibinfo{year}{1985}).

\bibitem[{\citenamefont{Kozlov et~al.}(1987)\citenamefont{Kozlov, Fomichev,
  Dmitriev, Labzovsky, and Titov}}]{Kozlov:87}
\bibinfo{author}{\bibfnamefont{M.~G.} \bibnamefont{Kozlov}},
  \bibinfo{author}{\bibfnamefont{V.~I.} \bibnamefont{Fomichev}},
  \bibinfo{author}{\bibfnamefont{{\hbox{Yu}}.~{\hbox{Yu}}.}
  \bibnamefont{Dmitriev}}, \bibinfo{author}{\bibfnamefont{L.~N.}
  \bibnamefont{Labzovsky}}, \bibnamefont{and}
  \bibinfo{author}{\bibfnamefont{A.~V.} \bibnamefont{Titov}},
  \bibinfo{journal}{J.\ Phys.\ B} \textbf{\bibinfo{volume}{20}},
  \bibinfo{pages}{4939} (\bibinfo{year}{1987}).

\bibitem[{\citenamefont{Sauer et~al.}(1994)\citenamefont{Sauer, Wang, and
  Hinds}}]{Sauer:94}
\bibinfo{author}{\bibfnamefont{B.~E.} \bibnamefont{Sauer}},
  \bibinfo{author}{\bibfnamefont{J.}~\bibnamefont{Wang}}, \bibnamefont{and}
  \bibinfo{author}{\bibfnamefont{E.~A.} \bibnamefont{Hinds}},
  \bibinfo{journal}{Bull.\ Am.\ Phys.\ Soc., Ser.~II}
  \textbf{\bibinfo{volume}{39}}, \bibinfo{pages}{1060} (\bibinfo{year}{1994}).

\bibitem[{\citenamefont{{Hudson} et~al.}(2011)\citenamefont{{Hudson}, {Kara},
  {Smallman}, {Sauer}, {Tarbutt}, and {Hinds}}}]{Hudson:11a}
\bibinfo{author}{\bibfnamefont{J.~J.} \bibnamefont{{Hudson}}},
  \bibinfo{author}{\bibfnamefont{D.~M.} \bibnamefont{{Kara}}},
  \bibinfo{author}{\bibfnamefont{I.~J.} \bibnamefont{{Smallman}}},
  \bibinfo{author}{\bibfnamefont{B.~E.} \bibnamefont{{Sauer}}},
  \bibinfo{author}{\bibfnamefont{M.~R.} \bibnamefont{{Tarbutt}}},
  \bibnamefont{and} \bibinfo{author}{\bibfnamefont{E.~A.}
  \bibnamefont{{Hinds}}}, \bibinfo{journal}{\nat}
  \textbf{\bibinfo{volume}{473}}, \bibinfo{pages}{493} (\bibinfo{year}{2011}).

\bibitem[{\citenamefont{Regan et~al.}(2002)\citenamefont{Regan, Commins,
  Schmidt, and DeMille}}]{Regan:02}
\bibinfo{author}{\bibfnamefont{B.~C.} \bibnamefont{Regan}},
  \bibinfo{author}{\bibfnamefont{E.~D.} \bibnamefont{Commins}},
  \bibinfo{author}{\bibfnamefont{C.~J.} \bibnamefont{Schmidt}},
  \bibnamefont{and} \bibinfo{author}{\bibfnamefont{D.}~\bibnamefont{DeMille}},
  \bibinfo{journal}{Phys.\ Rev.\ Lett.} \textbf{\bibinfo{volume}{88}},
  \bibinfo{pages}{071805} (\bibinfo{year}{2002}).

\bibitem[{\citenamefont{Baron et~al.}(2014)\citenamefont{Baron, Campbell,
  DeMille, Doyle, Gabrielse, Gurevich, Hess, Hutzler, Kirilov, Kozyryev
  et~al.}}]{ACME:14a}
\bibinfo{author}{\bibfnamefont{J.}~\bibnamefont{Baron}},
  \bibinfo{author}{\bibfnamefont{W.~C.} \bibnamefont{Campbell}},
  \bibinfo{author}{\bibfnamefont{D.}~\bibnamefont{DeMille}},
  \bibinfo{author}{\bibfnamefont{J.~M.} \bibnamefont{Doyle}},
  \bibinfo{author}{\bibfnamefont{G.}~\bibnamefont{Gabrielse}},
  \bibinfo{author}{\bibfnamefont{Y.~V.} \bibnamefont{Gurevich}},
  \bibinfo{author}{\bibfnamefont{P.~W.} \bibnamefont{Hess}},
  \bibinfo{author}{\bibfnamefont{N.~R.} \bibnamefont{Hutzler}},
  \bibinfo{author}{\bibfnamefont{E.}~\bibnamefont{Kirilov}},
  \bibinfo{author}{\bibfnamefont{I.}~\bibnamefont{Kozyryev}},
  \bibnamefont{et~al.} (\bibinfo{collaboration}{The ACME Collaboration}),
  \bibinfo{journal}{Science} \textbf{\bibinfo{volume}{343}},
  \bibinfo{pages}{269} (\bibinfo{year}{2014}).

\bibitem[{\citenamefont{Pospelov and Ritz}(2014)}]{Pospelov:14}
\bibinfo{author}{\bibfnamefont{M.}~\bibnamefont{Pospelov}} \bibnamefont{and}
  \bibinfo{author}{\bibfnamefont{A.}~\bibnamefont{Ritz}},
  \bibinfo{journal}{Phys.\ Rev.\ D} \textbf{\bibinfo{volume}{89}},
  \bibinfo{pages}{056006} (\bibinfo{year}{2014}).

\bibitem[{\citenamefont{Flambaum et~al.}(2014)\citenamefont{Flambaum, DeMille,
  and Kozlov}}]{FDK14}
\bibinfo{author}{\bibfnamefont{V.~V.} \bibnamefont{Flambaum}},
  \bibinfo{author}{\bibfnamefont{D.}~\bibnamefont{DeMille}}, \bibnamefont{and}
  \bibinfo{author}{\bibfnamefont{M.~G.} \bibnamefont{Kozlov}},
  \bibinfo{journal}{Phys.\ Rev.\ Lett.} \textbf{\bibinfo{volume}{113}},
  \bibinfo{pages}{103003} (\bibinfo{year}{2014}).

\bibitem[{\citenamefont{Skripnikov et~al.}(2014)\citenamefont{Skripnikov,
  Petrov, Titov, and Flambaum}}]{Skripnikov:14a}
\bibinfo{author}{\bibfnamefont{L.~V.} \bibnamefont{Skripnikov}},
  \bibinfo{author}{\bibfnamefont{A.~N.} \bibnamefont{Petrov}},
  \bibinfo{author}{\bibfnamefont{A.~V.} \bibnamefont{Titov}}, \bibnamefont{and}
  \bibinfo{author}{\bibfnamefont{V.~V.} \bibnamefont{Flambaum}}
  (\bibinfo{year}{2014}), \bibinfo{note}{arXiv:1408.5368 [physics.atom-ph]}.

\bibitem[{\citenamefont{Leanhardt et~al.}(2011)\citenamefont{Leanhardt, Bohn,
  Loh, Maletinsky, Meyer, Sinclair, Stutz, and Cornell}}]{Leanhardt:2011}
\bibinfo{author}{\bibfnamefont{A.}~\bibnamefont{Leanhardt}},
  \bibinfo{author}{\bibfnamefont{J.}~\bibnamefont{Bohn}},
  \bibinfo{author}{\bibfnamefont{H.}~\bibnamefont{Loh}},
  \bibinfo{author}{\bibfnamefont{P.}~\bibnamefont{Maletinsky}},
  \bibinfo{author}{\bibfnamefont{E.}~\bibnamefont{Meyer}},
  \bibinfo{author}{\bibfnamefont{L.}~\bibnamefont{Sinclair}},
  \bibinfo{author}{\bibfnamefont{R.}~\bibnamefont{Stutz}}, \bibnamefont{and}
  \bibinfo{author}{\bibfnamefont{E.}~\bibnamefont{Cornell}},
  \bibinfo{journal}{Journal of Molecular Spectroscopy}
  \textbf{\bibinfo{volume}{270}}, \bibinfo{pages}{1 } (\bibinfo{year}{2011}),
  ISSN \bibinfo{issn}{0022-2852}.

\bibitem[{\citenamefont{Skripnikov
  et~al.}(2013{\natexlab{a}})\citenamefont{Skripnikov, Petrov, and
  Titov}}]{Skripnikov:13c}
\bibinfo{author}{\bibfnamefont{L.~V.} \bibnamefont{Skripnikov}},
  \bibinfo{author}{\bibfnamefont{A.~N.} \bibnamefont{Petrov}},
  \bibnamefont{and} \bibinfo{author}{\bibfnamefont{A.~V.} \bibnamefont{Titov}},
  \bibinfo{journal}{J.\ Chem.\ Phys.} \textbf{\bibinfo{volume}{139}},
  \bibinfo{eid}{221103} (\bibinfo{year}{2013}{\natexlab{a}}).

\bibitem[{\citenamefont{Hutzler et~al.}(2012)\citenamefont{Hutzler, Hess,
  Kirilov, O'Leary, Petrik, Spaun, DeMille, Gabrielse, and
  Doyle}}]{ACME_Improvements}
\bibinfo{author}{\bibfnamefont{N.}~\bibnamefont{Hutzler}},
  \bibinfo{author}{\bibfnamefont{P.}~\bibnamefont{Hess}},
  \bibinfo{author}{\bibfnamefont{E.}~\bibnamefont{Kirilov}},
  \bibinfo{author}{\bibfnamefont{B.}~\bibnamefont{O'Leary}},
  \bibinfo{author}{\bibfnamefont{E.}~\bibnamefont{Petrik}},
  \bibinfo{author}{\bibfnamefont{B.}~\bibnamefont{Spaun}},
  \bibinfo{author}{\bibfnamefont{D.}~\bibnamefont{DeMille}},
  \bibinfo{author}{\bibfnamefont{G.}~\bibnamefont{Gabrielse}},
  \bibnamefont{and} \bibinfo{author}{\bibfnamefont{J.}~\bibnamefont{Doyle}},
  \bibinfo{journal}{Bull. Am. Phys. Soc.} \textbf{\bibinfo{volume}{57}},
  \bibinfo{pages}{H6.007} (\bibinfo{year}{2012}).

\bibitem[{\citenamefont{Petrov et~al.}(2014)\citenamefont{Petrov, Skripnikov,
  Titov, Hutzler, Hess, O'Leary, Spaun, DeMille, Gabrielse, and
  Doyle}}]{Petrov:14}
\bibinfo{author}{\bibfnamefont{A.~N.} \bibnamefont{Petrov}},
  \bibinfo{author}{\bibfnamefont{L.~V.} \bibnamefont{Skripnikov}},
  \bibinfo{author}{\bibfnamefont{A.~V.} \bibnamefont{Titov}},
  \bibinfo{author}{\bibfnamefont{N.~R.} \bibnamefont{Hutzler}},
  \bibinfo{author}{\bibfnamefont{P.~W.} \bibnamefont{Hess}},
  \bibinfo{author}{\bibfnamefont{B.~R.} \bibnamefont{O'Leary}},
  \bibinfo{author}{\bibfnamefont{B.}~\bibnamefont{Spaun}},
  \bibinfo{author}{\bibfnamefont{D.}~\bibnamefont{DeMille}},
  \bibinfo{author}{\bibfnamefont{G.}~\bibnamefont{Gabrielse}},
  \bibnamefont{and} \bibinfo{author}{\bibfnamefont{J.~M.} \bibnamefont{Doyle}},
  \bibinfo{journal}{Phys. Rev. A} \textbf{\bibinfo{volume}{89}},
  \bibinfo{pages}{062505} (\bibinfo{year}{2014}).

\bibitem[{\citenamefont{Fleig and Nayak}(2014)}]{Fleig:14}
\bibinfo{author}{\bibfnamefont{T.}~\bibnamefont{Fleig}} \bibnamefont{and}
  \bibinfo{author}{\bibfnamefont{M.~K.} \bibnamefont{Nayak}},
  \bibinfo{journal}{Journal of Molecular Spectroscopy}
  \textbf{\bibinfo{volume}{300}}, \bibinfo{pages}{16 } (\bibinfo{year}{2014}),
  ISSN \bibinfo{issn}{0022-2852}, \bibinfo{note}{spectroscopic Tests of
  Fundamental Physics}.

\bibitem[{\citenamefont{Titov et~al.}(2014)\citenamefont{Titov, Lomachuk, and
  Skripnikov}}]{Titov:14}
\bibinfo{author}{\bibfnamefont{A.~V.} \bibnamefont{Titov}},
  \bibinfo{author}{\bibfnamefont{Y.~V.} \bibnamefont{Lomachuk}},
  \bibnamefont{and} \bibinfo{author}{\bibfnamefont{L.~V.}
  \bibnamefont{Skripnikov}} (\bibinfo{year}{2014}),
  \bibinfo{note}{arXiv:1405.6892}.

\bibitem[{\citenamefont{Titov et~al.}(2006)\citenamefont{Titov, Mosyagin,
  Petrov, Isaev, and DeMille}}]{Titov:06amin}
\bibinfo{author}{\bibfnamefont{A.~V.} \bibnamefont{Titov}},
  \bibinfo{author}{\bibfnamefont{N.~S.} \bibnamefont{Mosyagin}},
  \bibinfo{author}{\bibfnamefont{A.~N.} \bibnamefont{Petrov}},
  \bibinfo{author}{\bibfnamefont{T.~A.} \bibnamefont{Isaev}}, \bibnamefont{and}
  \bibinfo{author}{\bibfnamefont{D.~P.} \bibnamefont{DeMille}},
  \bibinfo{journal}{Progr.\ Theor.\ Chem.\ Phys.}
  \textbf{\bibinfo{volume}{B~15}}, \bibinfo{pages}{253} (\bibinfo{year}{2006}).

\bibitem[{\citenamefont{Petrov et~al.}(2002)\citenamefont{Petrov, Mosyagin,
  Isaev, Titov, Ezhov, Eliav, and Kaldor}}]{Petrov:02}
\bibinfo{author}{\bibfnamefont{A.~N.} \bibnamefont{Petrov}},
  \bibinfo{author}{\bibfnamefont{N.~S.} \bibnamefont{Mosyagin}},
  \bibinfo{author}{\bibfnamefont{T.~A.} \bibnamefont{Isaev}},
  \bibinfo{author}{\bibfnamefont{A.~V.} \bibnamefont{Titov}},
  \bibinfo{author}{\bibfnamefont{V.~F.} \bibnamefont{Ezhov}},
  \bibinfo{author}{\bibfnamefont{E.}~\bibnamefont{Eliav}}, \bibnamefont{and}
  \bibinfo{author}{\bibfnamefont{U.}~\bibnamefont{Kaldor}},
  \bibinfo{journal}{Phys.\ Rev.\ Lett.} \textbf{\bibinfo{volume}{88}},
  \bibinfo{pages}{073001} (\bibinfo{year}{2002}).

\bibitem[{\citenamefont{Titov and Mosyagin}(1999)}]{Titov:99}
\bibinfo{author}{\bibfnamefont{A.~V.} \bibnamefont{Titov}} \bibnamefont{and}
  \bibinfo{author}{\bibfnamefont{N.~S.} \bibnamefont{Mosyagin}},
  \bibinfo{journal}{Int.\ J.\ Quantum Chem.} \textbf{\bibinfo{volume}{71}},
  \bibinfo{pages}{359} (\bibinfo{year}{1999}).

\bibitem[{\citenamefont{Mosyagin et~al.}(2010)\citenamefont{Mosyagin,
  Zaitsevskii, and Titov}}]{Mosyagin:10a}
\bibinfo{author}{\bibfnamefont{N.~S.} \bibnamefont{Mosyagin}},
  \bibinfo{author}{\bibfnamefont{A.~V.} \bibnamefont{Zaitsevskii}},
  \bibnamefont{and} \bibinfo{author}{\bibfnamefont{A.~V.} \bibnamefont{Titov}},
  \bibinfo{journal}{Review of Atomic and Molecular Physics}
  \textbf{\bibinfo{volume}{1}}, \bibinfo{pages}{63} (\bibinfo{year}{2010}).

\bibitem[{\citenamefont{Petrov et~al.}(2004)\citenamefont{Petrov, Mosyagin,
  Titov, and Tupitsyn}}]{Petrov:04b}
\bibinfo{author}{\bibfnamefont{A.~N.} \bibnamefont{Petrov}},
  \bibinfo{author}{\bibfnamefont{N.~S.} \bibnamefont{Mosyagin}},
  \bibinfo{author}{\bibfnamefont{A.~V.} \bibnamefont{Titov}}, \bibnamefont{and}
  \bibinfo{author}{\bibfnamefont{I.~I.} \bibnamefont{Tupitsyn}},
  \bibinfo{journal}{J.\ Phys.\ B} \textbf{\bibinfo{volume}{37}},
  \bibinfo{pages}{4621} (\bibinfo{year}{2004}).

\bibitem[{\citenamefont{Mosyagin et~al.}(2006)\citenamefont{Mosyagin, Petrov,
  Titov, and Tupitsyn}}]{Mosyagin:06amin}
\bibinfo{author}{\bibfnamefont{N.~S.} \bibnamefont{Mosyagin}},
  \bibinfo{author}{\bibfnamefont{A.~N.} \bibnamefont{Petrov}},
  \bibinfo{author}{\bibfnamefont{A.~V.} \bibnamefont{Titov}}, \bibnamefont{and}
  \bibinfo{author}{\bibfnamefont{I.~I.} \bibnamefont{Tupitsyn}},
  \bibinfo{journal}{Progr.\ Theor.\ Chem.\ Phys.}
  \textbf{\bibinfo{volume}{B~15}}, \bibinfo{pages}{229} (\bibinfo{year}{2006}).

\bibitem[{\citenamefont{Tupitsyn}(2003)}]{HFDB}
\bibinfo{author}{\bibfnamefont{I.~I.} \bibnamefont{Tupitsyn}}
  (\bibinfo{year}{2003}), \bibinfo{note}{{\sc hfdb}, a program for atomic
  finite-difference four-component {D}irac-{H}artree-{F}ock-{B}reit
  calculations written on the base of the {\sc hfd} code~\cite{Bratzev:77}}.

\bibitem[{\citenamefont{Bratzev et~al.}(1977)\citenamefont{Bratzev, Deyneka,
  and Tupitsyn}}]{Bratzev:77}
\bibinfo{author}{\bibfnamefont{V.~F.} \bibnamefont{Bratzev}},
  \bibinfo{author}{\bibfnamefont{G.~B.} \bibnamefont{Deyneka}},
  \bibnamefont{and} \bibinfo{author}{\bibfnamefont{I.~I.}
  \bibnamefont{Tupitsyn}}, \bibinfo{journal}{Bull.\ Acad.\ Sci.\ USSR, Phys.\
  Ser.} \textbf{\bibinfo{volume}{41}}, \bibinfo{pages}{173}
  (\bibinfo{year}{1977}).

\bibitem[{\citenamefont{Tupitsyn and Mosyagin}(1995)}]{HFJ}
\bibinfo{author}{\bibfnamefont{I.~I.} \bibnamefont{Tupitsyn}} \bibnamefont{and}
  \bibinfo{author}{\bibfnamefont{N.~S.} \bibnamefont{Mosyagin}}
  (\bibinfo{year}{1995}), \bibinfo{note}{{\sc hfj}, a program for atomic
  finite-difference two-component {H}artree-{F}ock calculations with the
  {g}eneralized {RECP} in the {$jj$}-coupling scheme}.

\bibitem[{\citenamefont{Tupitsyn et~al.}(1995)\citenamefont{Tupitsyn, Mosyagin,
  and Titov}}]{Tupitsyn:95}
\bibinfo{author}{\bibfnamefont{I.~I.} \bibnamefont{Tupitsyn}},
  \bibinfo{author}{\bibfnamefont{N.~S.} \bibnamefont{Mosyagin}},
  \bibnamefont{and} \bibinfo{author}{\bibfnamefont{A.~V.} \bibnamefont{Titov}},
  \bibinfo{journal}{J.\ Chem.\ Phys.} \textbf{\bibinfo{volume}{103}},
  \bibinfo{pages}{6548} (\bibinfo{year}{1995}).

\bibitem[{\citenamefont{Kozlov and Labzowsky}(1995)}]{Kozlov:95}
\bibinfo{author}{\bibfnamefont{M.}~\bibnamefont{Kozlov}} \bibnamefont{and}
  \bibinfo{author}{\bibfnamefont{L.}~\bibnamefont{Labzowsky}},
  \bibinfo{journal}{J.\ Phys.\ B} \textbf{\bibinfo{volume}{28}},
  \bibinfo{pages}{1933} (\bibinfo{year}{1995}).

\bibitem[{\citenamefont{Hunter}(1991)}]{Hunter:91}
\bibinfo{author}{\bibfnamefont{L.~R.} \bibnamefont{Hunter}},
  \bibinfo{journal}{Science} \textbf{\bibinfo{volume}{252}},
  \bibinfo{pages}{73} (\bibinfo{year}{1991}).

\bibitem[{\citenamefont{Kozlov et~al.}(1997)\citenamefont{Kozlov, Titov,
  Mosyagin, and Souchko}}]{Kozlov:97}
\bibinfo{author}{\bibfnamefont{M.~G.} \bibnamefont{Kozlov}},
  \bibinfo{author}{\bibfnamefont{A.~V.} \bibnamefont{Titov}},
  \bibinfo{author}{\bibfnamefont{N.~S.} \bibnamefont{Mosyagin}},
  \bibnamefont{and} \bibinfo{author}{\bibfnamefont{P.~V.}
  \bibnamefont{Souchko}}, \bibinfo{journal}{Phys.\ Rev.\ A}
  \textbf{\bibinfo{volume}{56}}, \bibinfo{pages}{R3326} (\bibinfo{year}{1997}).

\bibitem[{\citenamefont{Safronova et~al.}(2013)\citenamefont{Safronova,
  Safronova, Radnaev, Campbell, and Kuzmich}}]{Safronova:13}
\bibinfo{author}{\bibfnamefont{M.~S.} \bibnamefont{Safronova}},
  \bibinfo{author}{\bibfnamefont{U.~I.} \bibnamefont{Safronova}},
  \bibinfo{author}{\bibfnamefont{A.~G.} \bibnamefont{Radnaev}},
  \bibinfo{author}{\bibfnamefont{C.~J.} \bibnamefont{Campbell}},
  \bibnamefont{and} \bibinfo{author}{\bibfnamefont{A.}~\bibnamefont{Kuzmich}},
  \bibinfo{journal}{Phys. Rev. A} \textbf{\bibinfo{volume}{88}},
  \bibinfo{pages}{060501} (\bibinfo{year}{2013}).

\bibitem[{\citenamefont{Kirilov et~al.}(2013)\citenamefont{Kirilov, Campbell,
  Doyle, Gabrielse, Gurevich, Hess, Hutzler, O'Leary, Petrik, Spaun
  et~al.}}]{Kirilov:13}
\bibinfo{author}{\bibfnamefont{E.}~\bibnamefont{Kirilov}},
  \bibinfo{author}{\bibfnamefont{W.~C.} \bibnamefont{Campbell}},
  \bibinfo{author}{\bibfnamefont{J.~M.} \bibnamefont{Doyle}},
  \bibinfo{author}{\bibfnamefont{G.}~\bibnamefont{Gabrielse}},
  \bibinfo{author}{\bibfnamefont{Y.~V.} \bibnamefont{Gurevich}},
  \bibinfo{author}{\bibfnamefont{P.~W.} \bibnamefont{Hess}},
  \bibinfo{author}{\bibfnamefont{N.~R.} \bibnamefont{Hutzler}},
  \bibinfo{author}{\bibfnamefont{B.~R.} \bibnamefont{O'Leary}},
  \bibinfo{author}{\bibfnamefont{E.}~\bibnamefont{Petrik}},
  \bibinfo{author}{\bibfnamefont{B.}~\bibnamefont{Spaun}},
  \bibnamefont{et~al.}, \bibinfo{journal}{Phys.\ Rev.\ A}
  \textbf{\bibinfo{volume}{88}}, \bibinfo{pages}{013844}
  (\bibinfo{year}{2013}).

\bibitem[{\citenamefont{Wang et~al.}(2011)\citenamefont{Wang, Le, Steimle, and
  Heaven}}]{Wang:11c}
\bibinfo{author}{\bibfnamefont{F.}~\bibnamefont{Wang}},
  \bibinfo{author}{\bibfnamefont{A.}~\bibnamefont{Le}},
  \bibinfo{author}{\bibfnamefont{T.~C.} \bibnamefont{Steimle}},
  \bibnamefont{and} \bibinfo{author}{\bibfnamefont{M.~C.}
  \bibnamefont{Heaven}}, \bibinfo{journal}{J.\ Chem.\ Phys.}
  \textbf{\bibinfo{volume}{134}}, \bibinfo{pages}{031102}
  (\bibinfo{year}{2011}).

\bibitem[{\citenamefont{Buchachenko}(2010)}]{Buchachenko:10}
\bibinfo{author}{\bibfnamefont{A.~A.} \bibnamefont{Buchachenko}},
  \bibinfo{journal}{J.\ Chem.\ Phys.} \textbf{\bibinfo{volume}{133}},
  \bibinfo{pages}{041102} (\bibinfo{year}{2010}).

\bibitem[{\citenamefont{Paulovi\v{c} et~al.}(2003)\citenamefont{Paulovi\v{c},
  Nakajima, Hirao, Lindh, and Malmqvist}}]{Paulovic:03}
\bibinfo{author}{\bibfnamefont{J.}~\bibnamefont{Paulovi\v{c}}},
  \bibinfo{author}{\bibfnamefont{T.}~\bibnamefont{Nakajima}},
  \bibinfo{author}{\bibfnamefont{K.}~\bibnamefont{Hirao}},
  \bibinfo{author}{\bibfnamefont{R.}~\bibnamefont{Lindh}}, \bibnamefont{and}
  \bibinfo{author}{\bibfnamefont{P.~A.} \bibnamefont{Malmqvist}},
  \bibinfo{journal}{The Journal of Chemical Physics}
  \textbf{\bibinfo{volume}{119}}, \bibinfo{pages}{798} (\bibinfo{year}{2003}).

\bibitem[{\citenamefont{Meyer and Bohn}(2008)}]{Meyer:08}
\bibinfo{author}{\bibfnamefont{E.~R.} \bibnamefont{Meyer}} \bibnamefont{and}
  \bibinfo{author}{\bibfnamefont{J.~L.} \bibnamefont{Bohn}},
  \bibinfo{journal}{Phys.\ Rev.\ A} \textbf{\bibinfo{volume}{78}},
  \bibinfo{pages}{010502(R)} (\bibinfo{year}{2008}).

\bibitem[{DIR()}]{DIRAC12}
\bibinfo{note}{{DIRAC}, a relativistic ab initio electronic structure program,
  Release {DIRAC12} (2012), written by H.~J.~{\relax Aa}.~Jensen, R.~Bast,
  T.~Saue, and L.~Visscher, with contributions from V.~Bakken, K.~G.~Dyall,
  S.~Dubillard, U.~Ekstr{\"o}m, E.~Eliav, T.~Enevoldsen, T.~Fleig,
  O.~Fossgaard, A.~S.~P.~Gomes, T.~Helgaker, J.~K.~L{\ae}rdahl, Y.~S.~Lee,
  J.~Henriksson, M.~Ilia{\v{s}}, Ch.~R.~Jacob, S.~Knecht, S.~Komorovsk{\'y},
  O.~Kullie, C.~V.~Larsen, H.~S.~Nataraj, P.~Norman, G.~Olejniczak, J.~Olsen,
  Y.~C.~Park, J.~K.~Pedersen, M.~Pernpointner, K.~Ruud, P.~Sa{\l}ek,
  B.~Schimmelpfennig, J.~Sikkema, A.~J.~Thorvaldsen, J.~Thyssen,
  J.~van~Stralen, S.~Villaume, O.~Visser, T.~Winther, and S.~Yamamoto (see
  http://www.diracprogram.org)}.

\bibitem[{\citenamefont{Stanton et~al.}(2011)\citenamefont{Stanton, Gauss,
  Harding, Szalay et~al.}}]{CFOUR}
\bibinfo{author}{\bibfnamefont{J.~F.} \bibnamefont{Stanton}},
  \bibinfo{author}{\bibfnamefont{J.}~\bibnamefont{Gauss}},
  \bibinfo{author}{\bibfnamefont{M.~E.} \bibnamefont{Harding}},
  \bibinfo{author}{\bibfnamefont{P.~G.} \bibnamefont{Szalay}},
  \bibnamefont{et~al.} (\bibinfo{year}{2011}), \bibinfo{note}{{\sc cfour}: a
  program package for performing high-level quantum chemical calculations on
  atoms and molecules, {http://www.cfour.de} .}

\bibitem[{\citenamefont{Gauss et~al.}(1991)\citenamefont{Gauss, Lauderdale,
  Stanton, Watts, and Bartlett}}]{Gauss:91}
\bibinfo{author}{\bibfnamefont{J.}~\bibnamefont{Gauss}},
  \bibinfo{author}{\bibfnamefont{W.~J.} \bibnamefont{Lauderdale}},
  \bibinfo{author}{\bibfnamefont{J.~F.} \bibnamefont{Stanton}},
  \bibinfo{author}{\bibfnamefont{J.~D.} \bibnamefont{Watts}}, \bibnamefont{and}
  \bibinfo{author}{\bibfnamefont{R.~J.} \bibnamefont{Bartlett}},
  \bibinfo{journal}{Chem.\ Phys.\ Lett.} \textbf{\bibinfo{volume}{182}},
  \bibinfo{pages}{207 } (\bibinfo{year}{1991}).

\bibitem[{\citenamefont{Watts et~al.}(1993)\citenamefont{Watts, Gauss, and
  Bartlett}}]{Gauss:93}
\bibinfo{author}{\bibfnamefont{J.~D.} \bibnamefont{Watts}},
  \bibinfo{author}{\bibfnamefont{J.}~\bibnamefont{Gauss}}, \bibnamefont{and}
  \bibinfo{author}{\bibfnamefont{R.~J.} \bibnamefont{Bartlett}},
  \bibinfo{journal}{J.\ Chem.\ Phys.} \textbf{\bibinfo{volume}{98}},
  \bibinfo{pages}{8718} (\bibinfo{year}{1993}).

\bibitem[{MRC()}]{MRCC2013}
\bibinfo{note}{{\sc mrcc}, a quantum chemical program suite written by M.
  K\'{a}llay, Z. Rolik, I. Ladj\'{a}nszki, L. Szegedy, B. Lad\'{o}czki, J.
  Csontos, and B. Kornis. See also Z. Rolik and M. K\'{a}llay, J. Chem. Phys.
  135, 104111 (2011), as well as: www.mrcc.hu}.

\bibitem[{\citenamefont{Skripnikov and Titov}(2013)}]{Skripnikov:13b}
\bibinfo{author}{\bibfnamefont{L.~V.} \bibnamefont{Skripnikov}}
  \bibnamefont{and} \bibinfo{author}{\bibfnamefont{A.~V.} \bibnamefont{Titov}}
  (\bibinfo{year}{2013}), \bibinfo{note}{arXiv:1308.0163}.

\bibitem[{\citenamefont{Skripnikov et~al.}(2011)\citenamefont{Skripnikov,
  Titov, Petrov, Mosyagin, and Sushkov}}]{Skripnikov:11a}
\bibinfo{author}{\bibfnamefont{L.~V.} \bibnamefont{Skripnikov}},
  \bibinfo{author}{\bibfnamefont{A.~V.} \bibnamefont{Titov}},
  \bibinfo{author}{\bibfnamefont{A.~N.} \bibnamefont{Petrov}},
  \bibinfo{author}{\bibfnamefont{N.~S.} \bibnamefont{Mosyagin}},
  \bibnamefont{and} \bibinfo{author}{\bibfnamefont{O.~P.}
  \bibnamefont{Sushkov}}, \bibinfo{journal}{Phys.\ Rev.\ A}
  \textbf{\bibinfo{volume}{84}}, \bibinfo{pages}{022505}
  (\bibinfo{year}{2011}).

\bibitem[{\citenamefont{Kendall et~al.}(1992)\citenamefont{Kendall, {Dunning,
  Jr}, and Harrison}}]{Kendall:92}
\bibinfo{author}{\bibfnamefont{R.~A.} \bibnamefont{Kendall}},
  \bibinfo{author}{\bibfnamefont{T.~H.} \bibnamefont{{Dunning, Jr}}},
  \bibnamefont{and} \bibinfo{author}{\bibfnamefont{R.~J.}
  \bibnamefont{Harrison}}, \bibinfo{journal}{J.\ Chem.\ Phys.}
  \textbf{\bibinfo{volume}{96}}, \bibinfo{pages}{6796} (\bibinfo{year}{1992}).

\bibitem[{Note1()}]{Note1}
Note1, \bibinfo{note}{the number of $s$ basis functions was reduced because of
  a linear dependence problem, however, it did not influence the accuracy of
  the calculated parameters}.

\bibitem[{\citenamefont{Skripnikov
  et~al.}(2013{\natexlab{b}})\citenamefont{Skripnikov, Mosyagin, and
  Titov}}]{Skripnikov:13a}
\bibinfo{author}{\bibfnamefont{L.~V.} \bibnamefont{Skripnikov}},
  \bibinfo{author}{\bibfnamefont{N.~S.} \bibnamefont{Mosyagin}},
  \bibnamefont{and} \bibinfo{author}{\bibfnamefont{A.~V.} \bibnamefont{Titov}},
  \bibinfo{journal}{Chem.\ Phys.\ Lett.} \textbf{\bibinfo{volume}{555}},
  \bibinfo{pages}{79} (\bibinfo{year}{2013}{\natexlab{b}}).

\bibitem[{\citenamefont{{Alml\"of} and Taylor}(1987)}]{Almlof:87}
\bibinfo{author}{\bibfnamefont{J.}~\bibnamefont{{Alml\"of}}} \bibnamefont{and}
  \bibinfo{author}{\bibfnamefont{P.~R.} \bibnamefont{Taylor}},
  \bibinfo{journal}{J.\ Chem.\ Phys.} \textbf{\bibinfo{volume}{86}},
  \bibinfo{pages}{4070} (\bibinfo{year}{1987}).

\bibitem[{\citenamefont{Huber and Herzberg}(1979)}]{Huber:79}
\bibinfo{author}{\bibfnamefont{K.~P.} \bibnamefont{Huber}} \bibnamefont{and}
  \bibinfo{author}{\bibfnamefont{G.}~\bibnamefont{Herzberg}},
  \emph{\bibinfo{title}{Constants of Diatomic Molecules}}
  (\bibinfo{publisher}{Van Nostrand-Reinhold}, \bibinfo{address}{New York},
  \bibinfo{year}{1979}).

\bibitem[{\citenamefont{Edvinsson and Lagerqvist}(1984)}]{Edvinsson:84}
\bibinfo{author}{\bibfnamefont{G.}~\bibnamefont{Edvinsson}} \bibnamefont{and}
  \bibinfo{author}{\bibfnamefont{A.}~\bibnamefont{Lagerqvist}},
  \bibinfo{journal}{Physica Scripta} \textbf{\bibinfo{volume}{30}},
  \bibinfo{pages}{309} (\bibinfo{year}{1984}).

\bibitem[{\citenamefont{Bartlett}(1995)}]{Bartlett:95}
\bibinfo{author}{\bibfnamefont{R.~J.} \bibnamefont{Bartlett}}, in
  \emph{\bibinfo{booktitle}{Modern Electronic Stucture Theory}}, edited by
  \bibinfo{editor}{\bibfnamefont{D.~R.} \bibnamefont{Yarkony}}
  (\bibinfo{publisher}{World Scientific}, \bibinfo{address}{Singapore},
  \bibinfo{year}{1995}), vol.~\bibinfo{volume}{2} of
  \emph{\bibinfo{series}{Adv.\ Series in Phys.\ Chem.}}, pp.
  \bibinfo{pages}{1047--1131}, \bibinfo{note}{[Part~II]}.

\bibitem[{\citenamefont{Petrov et~al.}(2005)\citenamefont{Petrov, Titov, Isaev,
  Mosyagin, and DeMille}}]{Petrov:05a}
\bibinfo{author}{\bibfnamefont{A.~N.} \bibnamefont{Petrov}},
  \bibinfo{author}{\bibfnamefont{A.~V.} \bibnamefont{Titov}},
  \bibinfo{author}{\bibfnamefont{T.~A.} \bibnamefont{Isaev}},
  \bibinfo{author}{\bibfnamefont{N.~S.} \bibnamefont{Mosyagin}},
  \bibnamefont{and} \bibinfo{author}{\bibfnamefont{D.~P.}
  \bibnamefont{DeMille}}, \bibinfo{journal}{Phys.\ Rev.\ A}
  \textbf{\bibinfo{volume}{72}}, \bibinfo{pages}{022505}
  (\bibinfo{year}{2005}).

\bibitem[{\citenamefont{Isaev et~al.}(2004)\citenamefont{Isaev, Petrov,
  Mosyagin, Titov, Eliav, and Kaldor}}]{Isaev:04}
\bibinfo{author}{\bibfnamefont{T.~A.} \bibnamefont{Isaev}},
  \bibinfo{author}{\bibfnamefont{A.~N.} \bibnamefont{Petrov}},
  \bibinfo{author}{\bibfnamefont{N.~S.} \bibnamefont{Mosyagin}},
  \bibinfo{author}{\bibfnamefont{A.~V.} \bibnamefont{Titov}},
  \bibinfo{author}{\bibfnamefont{E.}~\bibnamefont{Eliav}}, \bibnamefont{and}
  \bibinfo{author}{\bibfnamefont{U.}~\bibnamefont{Kaldor}},
  \bibinfo{journal}{Phys.\ Rev.\ A} \textbf{\bibinfo{volume}{69}},
  \bibinfo{pages}{030501(R)} (\bibinfo{year}{2004}).

\bibitem[{\citenamefont{K\'{a}llay et~al.}(2002)\citenamefont{K\'{a}llay,
  Szalay, and Surj\'{a}n}}]{Kallay:2}
\bibinfo{author}{\bibfnamefont{M.}~\bibnamefont{K\'{a}llay}},
  \bibinfo{author}{\bibfnamefont{P.~G.} \bibnamefont{Szalay}},
  \bibnamefont{and} \bibinfo{author}{\bibfnamefont{P.~R.}
  \bibnamefont{Surj\'{a}n}}, \bibinfo{journal}{J.\ Chem.\ Phys.}
  \textbf{\bibinfo{volume}{117}}, \bibinfo{pages}{980} (\bibinfo{year}{2002}).

\bibitem[{\citenamefont{Vutha et~al.}(2011)\citenamefont{Vutha, Spaun,
  Gurevich, Hutzler, Kirilov, Doyle, Gabrielse, and DeMille}}]{Vutha:2011}
\bibinfo{author}{\bibfnamefont{A.~C.} \bibnamefont{Vutha}},
  \bibinfo{author}{\bibfnamefont{B.}~\bibnamefont{Spaun}},
  \bibinfo{author}{\bibfnamefont{Y.~V.} \bibnamefont{Gurevich}},
  \bibinfo{author}{\bibfnamefont{N.~R.} \bibnamefont{Hutzler}},
  \bibinfo{author}{\bibfnamefont{E.}~\bibnamefont{Kirilov}},
  \bibinfo{author}{\bibfnamefont{J.~M.} \bibnamefont{Doyle}},
  \bibinfo{author}{\bibfnamefont{G.}~\bibnamefont{Gabrielse}},
  \bibnamefont{and} \bibinfo{author}{\bibfnamefont{D.}~\bibnamefont{DeMille}},
  \bibinfo{journal}{Phys.\ Rev.\ A} \textbf{\bibinfo{volume}{84}},
  \bibinfo{pages}{034502} (\bibinfo{year}{2011}).

\bibitem[{Note2()}]{Note2}
Note2, \bibinfo{note}{due to computational/program limitations it was not
  possible to perform CCSDTQ-calculation in the two-component relativistic
  case}.

\end{thebibliography}

\end{document}